\documentclass[11pt]{article}
\usepackage{amsmath,amssymb,epsf,slashed}
\usepackage{color}

\makeatletter
\@addtoreset{equation}{section}
\makeatother

\def\ben{\begin{equation}}
\def\een{\end{equation}}

  \let\n=\nu  \let\p=\pi

\let\C=\Chi

\def\nn{\nonumber} \def\bd{\begin{document}} \def\ed{\end{document}}
\def\ds{\documentstyle} \let\fr=\frac \let\bl=\bigl \let\br=\bigr
\let\Br=\Bigr \let\Bl=\Bigl
\let\bm=\bibitem
\let\na=\nabla
\let\pa=\partial \let\ov=\overline
\newcommand{\be}{\begin{equation}}
\newcommand{\ee}{\end{equation}}
\def\ba{\begin{array}}
\def\ea{\end{array}}
\def\ft#1#2{{\textstyle{\frac{\scriptstyle #1}{\scriptstyle #2} } }}
\def\fft#1#2{{\frac{#1}{#2}}}
\def\del{\partial}
\def\vp{\varphi}
\def\sst#1{{\scriptscriptstyle #1}}
\def\oneone{\rlap 1\mkern4mu{\rm l}}
\def\td{\tilde}
\def\wtd{\widetilde}
\def\ie{{\it i.e.\ }}
\def\dalemb#1#2{{\vbox{\hrule height .#2pt
        \hbox{\vrule width.#2pt height#1pt \kern#1pt
                \vrule width.#2pt}
        \hrule height.#2pt}}}
\def\square{\mathord{\dalemb{6.8}{7}\hbox{\hskip1pt}}}
\newcommand{\ho}[1]{$\, ^{#1}$}
\newcommand{\hoch}[1]{$\, ^{#1}$}
\newcommand{\bea}{\begin{eqnarray}}
\newcommand{\eea}{\end{eqnarray}}
\newcommand{\ra}{\rightarrow}
\newcommand{\lra}{\longrightarrow}
\newcommand{\Lra}{\Leftrightarrow}
\newcommand{\bp}{\tilde \beta^\prime}
\newcommand{\tr}{{\rm tr} }
\newcommand{\Tr}{{\rm Tr} }
\def\0{{\sst{(0)}}}
\def\1{{\sst{(1)}}}
\def\2{{\sst{(2)}}}
\def\3{{\sst{(3)}}}
\def\4{{\sst{(4)}}}
\def\5{{\sst{(5)}}}
\def\6{{\sst{(6)}}}
\def\7{{\sst{(7)}}}
\def\8{{\sst{(8)}}}
\def\n{{\sst{(n)}}}
\def\cA{{{\cal A}}}
\def\cB{{{\cal B}}}
\def\cF{{{\cal F}}}
\def\cH{{{\cal H}}}
\def\tV{\widetilde V}
\def\tW{\widetilde W}
\def\tH{\widetilde H}
\def\tE{\widetilde E}
\def\tF{\widetilde F}
\def\tA{\widetilde A}
\def\im{{{\rm i}}}
\def\tY{{{\wtd Y}}}
\def\ep{{\epsilon}}
\def\vep{{\varepsilon}}
\def\bD{{{\bar D}}}
\def\R{{{\mathbb R}}}
\def\C{{{\mathbb C}}}
\def\H{{{\mathbb H}}}
\def\CP{{{\mathbb C}{\mathbb P}}}
\def\RP{{{\mathbb R}{\mathbb P}}}
\def\Z{{{\mathbb Z}}}
\def\bA{{{\mathbb A}}}
\def\bB{{{\mathbb B}}}
\def\bC{{{\mathbb C}}}
\def\bD{{{\mathbb D}}}
\def\bE{{{\mathbb E}}}
\def\bZ{{{\mathbb Z}}}
\def\Re{{{\frak{Re}}}}
\def\Im{{{\frak{Im}}}}
\def\cosec{{\,\hbox{cosec}\,}}
\def\Gm{{\Gamma_{\!\! -}}}
\def\Gp{{\Gamma_{\!\! +}}}
\def\stan{{standard }}
\def\nonstan{{supernumerary }}
\def\p{{\partial}}
\def\kdel#1{{\fft{\del}{\del#1}}}
\def\cN{{\cal N}}

\def\bog{{Bogomolny }}

\def\beps{{\bar\epsilon}}
\newcommand{\ww}[1]{\\[0.#1cm]}
\def\eps{\epsilon}
\def\slashchar#1{\setbox0=\hbox{$#1$}           
   \dimen0=\wd0                                 
   \setbox1=\hbox{/} \dimen1=\wd1               
   \ifdim\dimen0>\dimen1                        
      \rlap{\hbox to \dimen0{\hfil/\hfil}}      
      #1                                        
   \else                                        
      \rlap{\hbox to \dimen1{\hfil$#1$\hfil}}   
      /                                         
   \fi}
\def\sd{\slashchar{D}}
\def\sp{\slashchar{\partial}}
\newcommand{\eq}[1]{(\ref{#1})}
\def\cdh{\Gamma\cdot H}
\def\cL{e^{-1} {\cal L}}
\def\nb{{\bar\nabla}}
\def\nB{{\bar\Box}}
\def\th{{\widetilde h}}
\def\gb{{\bar g}}
\newcommand{\cD}{{\cal D}}

\begin{document}

\begin{flushright}
\hfill{ \
CAS-KITPC/ITP-193\ \ \ \ MIFPA-10-28\ \ \ \ }
\end{flushright}

\vspace{25pt}
\begin{center}
{\Large {\bf Massive Three-Dimensional
Supergravity From $R+R^2$ Action in Six Dimensions }}

\vspace{15pt}

{\Large H. L\"u\hoch{1,2}, C.N. Pope\hoch{3,4,5} and E. Sezgin\hoch{3,5}}

\vspace{20pt}

\hoch{1}{\it China Economics and Management Academy\\
Central University of Finance and Economics, Beijing 100081}

\vspace{10pt}

\hoch{2}{\it Institute for Advanced Study, Shenzhen
University, Nanhai Ave 3688, Shenzhen 518060}

\vspace{10pt}

\hoch{3} {\it George P. \& Cynthia Woods Mitchell  Institute
for Fundamental Physics and Astronomy,\\
Texas A\&M University, College Station, TX 77843, USA}

\vspace{10pt}

\hoch{4}{\it  DAMTP, Centre for Mathematical Sciences,
 Cambridge University,\\  Wilberforce Road, Cambridge CB3 OWA, UK}

\vspace{10pt}

\hoch{5}{\it Kavli Institute for Theoretical Physics China, CAS, Beijing
100190, China}

\vspace{40pt}

\underline{ABSTRACT}
\end{center}

  We obtain a three-parameter family of massive $\cN=1$  supergravities in
three dimensions from the 3-sphere reduction of an off-shell $\cN=(1,0)$
six-dimensional Poincar\'e supergravity that includes a curvature
squared invariant.  The three-dimensional theory contains an off-shell
supergravity multiplet and an on-shell scalar matter multiplet.  We then
generalise this in three dimensions to an eight-parameter family of
supergravities.  We also find a duality relationship between the
six-dimensional theory and the ${\cal N}=(1,0)$ six-dimensional theory
obtained through a $T^4$ reduction of the heterotic string effective
action that includes the higher-order terms associated with the
supersymmetrisation of the anomaly-cancelling $\tr(R\wedge R)$ term.

\vspace{15pt}

\thispagestyle{empty}

\newpage

\tableofcontents


\newpage


\section{Introduction}


  Three-dimensional topologically massive gravity (TMG) was
introduced long ago as a toy model for studying quantum gravity
\cite{desjactem,deskay}.  The topological mass term was
introduced in order to give dynamics to the gravitational field, but in
order for the graviton to have positive mass it was necessary to reverse
the usual sign of the Einstein-Hilbert term.  This had the unfortunate
effect that the BTZ black hole solution then has negative mass.
Recently, it was observed that if the coupling constant of the
topological mass term takes a certain critical value, the bulk massive
graviton decouples, one can revert to the usual sign for the
Einstein-Hilbert term, and thus one
obtains a unitary theory where the BTZ black hole has positive mass
\cite{lisost}. The theory is dual to a two-dimensional chiral
conformal field theory on the boundary.

   In subsequent developments in three-dimensional gravity,
further generalisations were introduced involving the addition of
higher-order curvature terms \cite{berhohtow,berhohtow2}.
These multi-parameter theories allow families of parameter choices
that again exhibit critical behaviour, with the massive graviton
decoupling in the bulk. Supersymmetric extensions of these
theories \cite{anbedrhoseto,behoroseto}, and also the original TMG
theory \cite{deskay,deser2,seztan,anbedrhoseto}, have been constructed.

   Even though the original motivation for considering three-dimensional
gravity theories was that they might stand in their own right as
consistent toy models for quantum gravity, it is nevertheless
natural to ask whether these theories can be embedded in string theory.
If this could be done, it could, as a toy model, help to shed light on
the quantum description of gravity within string theory in higher
dimensions. A first step was taken in \cite{gupsen}, where the
2-sphere reduction of a five-dimensional supergravity with
curvature-squared terms \cite{hanohatac} was considered.  After the
reduction in \cite{gupsen},
non-local and higher-derivative field redefinitions were
performed in order to remove the resulting curvature-squared terms in
three dimensions, giving precisely the usual TMG theory.  The use of such
field redefinitions is somewhat questionable, since the solution space of
the redefined theory is different from that of the original one
\cite{gupsen}.\footnote{By the same token, such non-local higher-derivative
field redefinitions could ostensibly be used to transform the ``new
massive gravity'' theory of \cite{berhohtow} into pure Einstein-Hilbert
gravity. Again, the non-local, and non-invertible, nature of the
transformations implies that the solution spaces in the two sets of
variables are inequivalent.} The string theory origin of the
five-dimensional theory, which is itself complete without requiring any
terms beyond the quadratic order in curvature, is unclear.

  More recently, it was shown that theories of gravity with topological
terms (of the form $\tr(R\wedge R\wedge\cdots R\wedge \omega)$, etc.) in
$4k+3$ dimensions can be recursively embedded in higher-dimensional
such theories via consistent reductions on homogeneous Einstein spaces of
dimension $4n$ \cite{lupa}.  A special case of such reductions
yields three-dimensional TMG from a
seven-dimensional starting point \cite{luwa}.

In this paper, we shall study the 3-sphere reduction of a six-dimensional
off-shell Poincar\'e supergravity with $R+ \alpha\, {\rm Riem}^2$ type
action, where $\alpha$ is an arbitrary constant, which was constructed in
\cite{Bergshoeff:1985mz,Bergshoeff:1986wc}. We find that this reduction,
upon a suitable and consistent truncation, yields a class of massive
$\cN=1$ supergravities in three dimensions which include a Lorentz
Chern-Simons term. These are disjoint from a larger class of such models
that were recently constructed directly in three dimensions
\cite{behoroseto}. We shall not carry out any non-local field
redefinitions of the kind described in \cite{gupsen}, which could
ostensibly allow us to remove all the higher-order curvature terms except
the Lorentz Chern-Simons term, for the reasons we discussed above.

In section 2, we present the action, supersymmetry transformation rules
and bosonic equations of motion for the six-dimensional
$R+\alpha\, {\rm Riem}^2$ supergravity model constructed in
\cite{Bergshoeff:1985mz,Bergshoeff:1986wc}.
In section 3, we perform a consistent reduction of the
six-dimensional theory on a 3-sphere, and obtain a three-parameter family
of massive supergravities in three dimensions.  (There are two continuous
parameters and a discrete parameter, which is the coefficient of the
Einstein-Hilbert term, and which takes the values 1, 0 or $-1$.)
These theories comprise
a mixing of an off-shell ${\cal N}= 1$ supergravity multiplet coupled
to an on-shell scalar multiplet.  We give the full
supersymmetric completion of the bosonic action, including those terms
which have not previously been studied in \cite{behoroseto}.
In section 4 we generalise our three-dimensional theory to one containing
eight parameters, thus extending previous results in \cite{behoroseto}.
We then study the critical points of these three-dimensional theories,
\ie the hypersurfaces in parameter space where the propagating
gravity modes decouple.

   Having sucessfully embedded a generalised three-dimensional
topologically massive supergravity in six dimensions, we can address the
further question of whether the embedding can be lifted to string theory.
The off-shell nature of the model ensures that the $R$ and ${\rm Riem}^2$
actions are separately supersymmetric. By contrast, adding higher derivative
terms in the on-shell supergravities, such as the ten-dimensional heterotic
supergravity, requires a derivative expansion which is supersymmetric
only order by order in $\alpha'$. In attempting to embed
the six-dimensional model we are studying here into ten-dimensional
heterotic supergravity, we find, as discussed in section 5,
that the 4-torus reduction of the latter,
followed by a consistent truncation, is related to the six-dimensional
model in a dual formulation.  More precisely, it appears that the
$T^4$ dimensional reduction of the (infinite number of)
 terms in the heterotic $\alpha'$
expansion that are directly associated with the supersymmetric
completion of the $\tr(R\wedge R)$ anomaly-cancelling term seem to be
reproduced simply by dualising the 2-form
potential of the exact $R+\alpha\, {\rm Riem}^2$ model.
We provide evidence for this
phenomenon by computing the leading terms in the correspondence between
these two models connected by duality transformation.

   Our conclusions appear in section 6.  Appendix A
sets out our notation and conventions, and appendix B contains some
formulae useful for the calculations in section 4.  In appendix C,
we present the six-dimensional off-shell $\cN=(1,0)$ supergravity model,
which was given in the string frame in section 2, in the Einstein frame.


\section{The Six-Dimensional Theory}



\subsection{The off-shell Poincar\'e multiplet and field redefinitions}


 Our starting point is the six-dimensional $\cN=(1,0)$ supergravity constructed
in \cite{Bergshoeff:1985mz}, using the $48+48$ component
off-shell Poincar\'e supermultiplet consisting of
the fields
\begin{equation}
\left\{ e_\mu^a, \psi_\mu^i, V_\mu^{ij},
B_{\mu\nu}, C_{\mu\nu\rho\sigma}, \phi, \psi^i\ \right\}\ ,
\end{equation}
where $e_\mu^a$ is the vielbein, $\psi_\mu^i$,  $(i=1,2)$, is the symplectic
Majorana-Weyl gravitino, $V_\mu^{ij}=V_\mu^{ji}$ is a triplet of $SU(2)$ gauge
fields, $B_{\mu\nu}$ is a real 2-form potential with 1-form gauge
invariance, $C_{\mu\nu\rho\sigma}$ is a real 4-form potential with
3-form gauge invariance, $\phi$ is a real scalar, and $\psi^i$ is a
symplectic Majorana-Weyl spinor. We shall work with the field strengths
$H=dB$ and $G=dC$.

The full Lagrangian we shall study is a sum of the off-shell Poincar\'e supergravity
\cite{Bergshoeff:1985mz} and an off-shell supersymmetrisation of $R_{\mu\nu ab}^2$ \cite{Bergshoeff:1986wc}. The sum can be schematically written as
\begin{equation}\label{twoactions}
\cL = \sigma R(e)+\cdots +\ft14 \alpha e^\phi \left( R_{\mu\nu ab}(e)^2+\cdots \right)\ ,
\end{equation}
where $\alpha$ is an constant, $\sigma=\pm 1, 0$, and we have set the
six-dimensional gravitational coupling constant to unity.
It was shown in \cite{Bergshoeff:1986wc}
that the field redefinitions\footnote{The redefinition of
$V_\mu^{ij}$ given in \cite{Bergshoeff:1986wc} is quadratic in fermions,
which is not relevant at the  order in fermions we are working in this paper.}
\begin{eqnarray}\label{redefs}
{\widehat e}_\mu^a &=& e^{\phi/2} e_\mu^a\ ,
\nn\ww2
{\widehat\psi}_\mu  &=& e^{\phi/4} \psi_\mu - e^{-3\phi/4} \Gamma_\mu\psi\ ,
\nn\ww2
{\widehat\eps} &=& e^{\phi/4}\eps \ .
\end{eqnarray}
lead to considerable simplifications in the supersymmetry transformation
rules (detailed in the next subsection and appendix C) that facilitates
the construction of the $R_{\mu\nu ab} R^{\mu\nu ab}$ invariant by analogy
with the super Yang-Mills system. In particular, the dilaton dependence
disappears in this invariant and the result now takes the form
\begin{equation}\label{rescaled}
\cL =  \sigma e^{-2\phi}\, R({\hat e})+\cdots +
\ft14 \alpha \left( R_{\mu\nu ab}({\hat e})^2+\cdots \right)
= e^{-1} \left( \sigma {\cal L}_R + \ft14  \alpha  {\cal L}_{Riem^2}\right)\ .
\end{equation}
%


\subsection{The bosonic part of the off-shell Poincar\'e action and
supersymmetry}


In more detail, and dropping the hats for simplicity, the bosonic
part of the full action is given by
\begin{eqnarray}
e^{-1} {\cal L}_R &=& e^{-2\phi}
\left( R +4\partial_\mu\phi \partial^\mu\phi
 -\ft{1}{12} H_{\mu\nu\rho} H^{\mu\nu\rho}
 +\ft12 V_\mu^{ij} V^\mu_{ij}\right)
\nn\ww2
&&  -\frac{1}{5!{\sqrt 2}} \varepsilon^{\mu\nu_1...\nu_5}
G_{\nu_1...\nu_5}V_{\mu ij}\delta^{ij}
+\frac{1}{2\times 5!} e^{2\phi} G_{\mu_1...\mu_5} G^{\mu_1...\mu_5}\ ,
\end{eqnarray}
and
\begin{equation}
e^{-1} {\cal L}_{Riem^2} =
R_{\mu\nu ab}(\omega_+) R^{\mu\nu ab} (\omega_+)
 - G_{\mu\nu}^{ij} G^{\mu\nu}_{ij}
+\ft14 \varepsilon^{\mu\nu\rho\sigma\lambda\tau}
R_{\mu\nu ab}(\omega_+) R_{\rho\sigma}{}^{ab}(\omega_+) B_{\lambda\tau}\ ,
\end{equation}
where $H_{\mu\nu\rho} e^\nu_a e^\rho_b =
H_{\mu ab}$ enters the spin connection as a torsion,
\begin{equation}\label{tc1}
\omega_{\mu\pm}{}^{ab} = \omega_\mu{}^{ab} \pm  \ft12 H_\mu{}^{ab}\ .
\end{equation}
The spin connection $\omega_{\mu ab}$ is the standard one that follows from
$de^a +\omega^a{}_b \wedge e^b =0$. The curvatures are defined as
\begin{eqnarray}
R_{\mu\nu}{}^{ab}(\omega_\pm) &=&
2\partial_{[\mu} \omega_{\nu]\pm}{}^{ab} +
\omega_{[\mu\pm}{}^{ac} \, \omega_{\nu]\pm}{}^{cb}\ ,
\nn\ww2
G_{\mu\nu}^{ij} &=& 2\partial_{[\mu} V_{\nu]}^{ij} +
 V_{[\mu}{}^{k(i} V_{\nu]}{}^{j)}{}_k\ .
\end{eqnarray}

The off-shell supersymmetry transformations, up to cubic fermion terms,
are given
by \cite{Bergshoeff:1985mz,Bergshoeff:1986wc} (see appendix C for
further details)\footnote{ In the results of \cite{Bergshoeff:1986wc}, we have
multiplied the action by  $2$ and let
$H_{\mu\nu\rho} \rightarrow -H_{\mu\nu\rho}$.
Furthermore, in eqn. (4.29) of \cite{Bergshoeff:1985mz}, $1/8 \to 1/4$ and
$-3/4\to -3/8$, thereby correcting the typos there. See appendix A for further
information on conventions.}

%
%
%
\begin{eqnarray}\label{susy6}
\delta e_\mu^a &=& \ft12 \beps \, \Gamma^a\psi_\mu\ ,
\nn\ww2
\delta \psi_\mu &=& {\cal D}_\mu (\omega_-) \eps\ ,
\nn\ww2
\delta B_{\mu\nu} &=& \beps \,\Gamma_{[\mu}\psi_{\nu]}\ ,
\nn\ww2
\delta V_\mu^{ij} &=&  \beps^{(i} \Gamma^\lambda \psi^{j)}_{\lambda\mu}(\omega_-)
 +\ft16 \beps^{(i} \Gamma\cdot H \psi^{j)}_\mu\ ,
\nn\ww2
\delta C_{\mu\nu\rho\sigma} &=& 2\sqrt2
e^{-2\phi} \beps^i\,\Gamma_{[\mu\nu\rho}  \psi^j_{\sigma]}\, \delta_{ij}+
2{\sqrt 2}  e^{-13\phi/4} \beps^i\, \Gamma_{\mu\nu\rho\sigma}\psi^j
\, \delta_{ij}\ ,
\nn\ww2
\delta\psi &=& \ft14 e^{5\phi/4}\Gamma^\mu \partial_\mu\phi\eps
-\ft{1}{48} e^{5\phi/4} \Gamma\cdot H  \eps-e^\phi \eta
\nn\ww2
\delta\phi &=& e^{-5\phi/4} \beps\psi\ ,
\end{eqnarray}
where $\Gamma\cdot H = \Gamma^{\mu\nu\rho}
H_{\mu\nu\rho}$.
The parameter $\eta$, up to cubic fermions, is defined by
\begin{equation}\label{eta3}
\eta_i =
\Big( -\frac1{4\sqrt2}  e^{\phi/4} \Gamma^\mu V_\mu^{(j}{}_k\,
 \delta^{\ell)k}\,\epsilon_j +
\frac{1}{5!\times 8{\sqrt 2}}
 e^{9\phi/4} \Gamma^{\mu_1...\mu_5} G_{\mu_1...\mu_5}\eps^\ell\Big)
\delta_{\ell i}\ ,
\end{equation}
and
\begin{eqnarray}
{\cal D}_\mu (\omega_-) \eps^i &=&\partial_\mu \eps^i +
\ft14\omega_{\mu ab}(\omega_-) \Gamma^{ab} \eps^i  +
\ft12 V_\mu^{ij} \eps_j\ ,
\nn\ww2
\psi^i_{\mu\nu}(\omega_-) &=& 2{\cal D}_{[\mu}(\omega_-) \psi^i_{\nu]}= \left(\partial_\mu +\ft14 \omega_{\mu-}{}^{ab}\Gamma_{ab}\right) \psi^i_\nu  +\ft12 V_\mu^{ij}\psi_{\nu j} - \ \  \mu \leftrightarrow \nu \ .
\label{defs2}
\end{eqnarray}
The transformation rules for $C_{\mu\nu\rho\sigma}$ and $\psi,\phi$
in the form above were not given in \cite{Bergshoeff:1986wc} because
they were not needed for the construction of the $R_{\mu\nu}{}^{ab}
R^{\mu\nu}{}_{ab}$
invariant which involves only the fields
$(e_\mu^a, \psi_\mu, B_{\mu\nu}, V_\mu^i)$.
We shall need these transformation rules, however, in the off-shell
Poincar\'e supergravity
sector.


\subsection{The $R_{\mu\nu ab}R^{\mu\nu ab}$ invariant}


The supersymmetrization of $R_{\mu\nu ab}R^{\mu\nu ab}$ in six dimensions
has been
accomplished in \cite{Bergshoeff:1986wc} by using the
Noether procedure, and in \cite{Bergshoeff:1987rb} by
exploiting a  map between the Yang-Mills supermultiplet and a set of
fields in the
off-shell Poincar\'e supermultiplet in \cite{Bergshoeff:1987rb}:
\begin{equation}\label{rules}
\left(A_\mu^I,\, Y^{ij}_I,\, \lambda^i_I\right) \quad\quad
\longrightarrow \quad\quad
\left(2\widehat\omega_\mu{}^{ab}_+,\, -
\widehat G_{ab}{}^{ij},\, \psi_{ab}^i(\omega_-)\right)\ ,
\end{equation}
where
\begin{equation}
{\widehat G}_{\mu\nu}^{ij} = G_{\mu\nu}^{ij}+2{\bar\psi}^{(i}_{[\mu}
\Gamma^\lambda \psi^{j)}_{\nu]\lambda}(\omega_-)
+\ft16 {\bar\psi}^{(i}_{[\mu}\Gamma\cdot H\psi_{\nu]}^{j)}\ ,
\end{equation}
up to quartic fermions, and
\begin{equation}\label{omegahat}
{\widehat\omega}_{\mu+}^{ab} =
\omega_\mu{}^{ab} +\ft12{\bar\psi}_\mu\Gamma^{[a}\psi^{b]}
  +\ft14 {\bar\psi}^a\Gamma_\mu\psi^b +\ft12 H_\mu{}^{ab}\ .
\end{equation}
As explained in \cite{Bergshoeff:1987rb}, applying this map
to the super Yang-Mills multiplet coupled to Poincar\'e supergravity given in
\cite{Bergshoeff:1985mz} leads to the action
\begin{eqnarray}\label{riem2}
\cL_{Riem^2} &=& R_{\mu\nu ab}(\widehat \omega_+)
R^{\mu\nu ab}(\widehat \omega_+) +2{\bar\psi}^{ab}(\omega_+)
\slashed \cD(\omega,\omega_+) \psi_{ab}(\omega_-)-
{\widehat G_{\mu\nu}}^{ij} {\widehat G^{\mu\nu}_{ij}}
\nn\ww2
&& -R^{\mu\nu ab}(\omega_-) {\bar\psi}_\lambda \Gamma_{ab}
\Gamma^\lambda\psi_{\mu\nu}(\omega_-)
+\ft1{12} {\bar\psi}^{ab}(\omega_-) \Gamma\cdot H \psi_{ab}(\omega_-)
\nn\ww2
&& +\ft14 \varepsilon^{\mu\nu\rho\sigma\lambda\tau}
R_{\mu\nu}{}^{ab}(\widehat\omega_+)
R_{\rho\sigma ab}(\widehat\omega_+) B_{\lambda\tau}\ ,
\end{eqnarray}
where, up to quartic fermions,
\begin{equation}\label{sdef}
\cD_\mu(\omega,\omega_+)\psi^i_{ab}=\left( \partial_\mu +
\ft14\omega_\mu{}^{cd}\Gamma_{cd}\right)\psi^i_{ab}
-2\omega_{\mu+[a}{}^c \psi^i_{b]c} + \ft12 V_{[a}^{ij} \,\psi_{b]j} \\ .
\end{equation}
The action is given modulo quartic fermions, and so in
addition to the quartic
fermion terms that have been suppressed, those which arise from the
term $R_{\mu\nu ab}(\hat\omega_+)R^{\mu\nu ab}(\hat\omega_+)$
and ${\widehat G}_{\mu\nu}^{ij}{\widehat G}^{\mu\nu}_{ij}$ can be dropped.
Note in particular that there will be terms bilinear in
fermions coming from the fermionic torsion
in the first term in the action. We have also used the fact that
$R_{\mu\nu a b}(\omega_-)=R_{ab\mu\nu}(\omega_+) +
{\rm fermion\ bilinears}$ \cite{Bergshoeff:1986wc}.

It will prove to be convenient to rewrite
${\bar\psi}^{ab} \slashed \cD (\omega,\omega_+)\psi_{ab}$  in terms of the torsionful
spin connection $\omega_{\mu-}{}^{ab}$. This leads to the action
\cite{Bergshoeff:1986wc,Bergshoeff:1987rb}
\begin{eqnarray}
\cL_{Riem^2} &=& R_{\mu\nu ab}(\widehat \omega_+)
R^{\mu\nu ab}(\widehat \omega_+) +2{\bar\psi}^{ab}(\omega_-)
\slashed \cD(\omega_-) \psi_{ab}(\omega_-)
-{\widehat G}_{\mu\nu}^{ij} {\widehat G}^{\mu\nu}_{ij}
\nn\ww2
&& -R^{\mu\nu ab}(\omega_-) {\bar\psi}_\lambda \Gamma_{ab}
\Gamma^\lambda\psi_{\mu\nu}(\omega_-)
+\ft13 {\bar\psi}^{ab}(\omega_-) \Gamma\cdot H \psi_{ab}(\omega_-)
\label{riem3}\ww2
&&+4 H^{\mu\nu\rho}{\bar\psi}_{\mu\lambda}(\omega_-)\Gamma_\rho \psi_\nu{}^\lambda (\omega_-)
+\ft14 \varepsilon^{\mu\nu\rho\sigma\lambda\tau} R_{\mu\nu}{}^{ab}(\omega_+)
R_{\rho\sigma ab}(\omega_+) B_{\lambda\tau}\ ,
\nn
\end{eqnarray}
where
\begin{equation}\label{sdef2}
\cD_\mu(\omega_-)\psi^i_{ab}=\left( \partial_\mu
  +\ft14\omega_{\mu-}{}^{cd}\Gamma_{cd}\right)\psi^i_{ab}
  -2\omega_{\mu-[a}{}^c \psi^i_{b])c} + \ft12 V_\mu^{ij} \,\psi_{ab j} \ .
\end{equation}
A few notational clarifications are in order.
Firstly, $\psi_{ab}(\omega_-)=e_a^\mu e_b^\nu\, \psi_{\mu\nu}(\omega_-)$,
with $\psi_{\mu\nu}(\omega_-)$ as defined in \eq{defs2}. Using
this definition in the second term in the above action gives
$2{\bar\psi}^{\mu\nu}(\omega_-)\slashed \cD(\omega_-,\Gamma_-)
\psi_{\mu\nu}(\omega_-)$
where
\begin{equation}\label{tc}
\Gamma_\pm^\rho{}_{\mu\nu} =
\Gamma^\rho{}_{\mu\nu} \pm \ft12 H^\rho{}_{\mu\nu}\ ,
\end{equation}
satisfying  the vielbein postulate
\begin{equation}\label{sp}
\partial_\mu e_\nu^a +\omega_{\mu\pm}{}^{ab} e_{\nu b}
-\Gamma_\mp^\rho{}_{\mu\nu}\, e_\rho^a = 0\ .
\end{equation}
and
\begin{equation}\label{cd5}
\cD_\lambda (\omega_-,\Gamma_-) \psi^i_{\mu\nu} = \left( \partial_\lambda +
\ft14\omega_{\lambda -}{}^{ab}\Gamma_{ab}\right)\psi^i_{\mu\nu}
+2\Gamma_-^\sigma{}_{\lambda[\mu}  \psi^i_{\nu]\sigma} + \ft12 V_\lambda^{ij} \,\psi_{\mu\nu j} \ .
\end{equation}


\subsection{The bosonic equations of motion}


After performing a consistent truncation of the auxiliary fields,
by setting
\begin{equation}\label{tr1}
V_\mu^{ij}=0\ , \quad\quad  C_{\mu\nu\rho\sigma}=0\ ,
\end{equation}
the six-dimensional action is given by
\begin{eqnarray}
{\cal L} &=&\sqrt{-g}\, e^{-2\phi}\,
 \Big[
 R + 4 (\del\phi)^2 - \ft1{12} H^{\mu\nu\rho} H_{\mu\nu\rho} \Big]
 +\ft14 \alpha \sqrt{-g}\,
\widetilde R^{\mu\nu\rho\sigma} \widetilde R_{\mu\nu\rho\sigma}
 +\ft14 \beta{\cal L}_{CS} \,,\label{sezginlag}
\end{eqnarray}
where $H=dB$ and
\begin{equation}
{\cal L}_{CS}=
\ft14 \epsilon^{\mu\nu\rho\sigma\lambda\tau}\,
\widetilde R^{\alpha\beta}{}_{\mu\nu}\, \widetilde R_{\alpha\beta\rho\sigma}\,
  B_{\lambda\tau}\,.
\end{equation}
Here we have introduced the additional parameter $\beta$ for convenience,
so that we can distinguish between terms coming from curvature squared,
versus terms coming from the Chern-Simons term. In what follows we should
keep in mind that six-dimensional supersymmetry will require
$\beta=\pm\alpha$, and in fact with the fermion conventions we shall be
using, supersymmetry requires
\begin{equation}\label{ab}
\beta=+\alpha\ .
\end{equation}
(For simplicity, we have not included the discrete parameter $\sigma$,
introduced in (\ref{twoactions}) and (\ref{rescaled}) here.  It can easily
be introduced if desired by making the field redefinition
$\phi\longrightarrow \phi -\ft12 \log\sigma$, with $\sigma$ initially
allowed to be any constant and then taken to be $\pm1$ or 0 after the
substitution.)

In the remainder of this section, we shall use the notation
\begin{equation}\label{cminus}
\Gamma_-^\rho{}_{\mu\nu} \equiv {\widetilde\Gamma}^\rho{}_{\mu\nu}=
 \Gamma^\rho{}_{\mu\nu} - \ft12 H^\rho{}_{\mu\nu}\ .
\end{equation}
The Riemann tensor for this connection
is defined by\footnote{From the vielbein postulate \eq{sp}, it follows that
$R_{\mu\nu}{}^{ab} (\omega_+)
e^\alpha_a e_{\beta b} = {\widetilde R}^\alpha{}_{\beta\mu\nu}$. Note that
when writing the Riemann tensor with all coordinate indices, we follow
the original and standard general relativity convention of putting the
manifestly-antisymmetric index pair to the right,
$R^\mu{}_{\nu\rho\sigma}=2\del_{[\rho}\Gamma^\mu{}_{\sigma]\nu} +\cdots$,
whereas when the Riemann
tensor is written with two coordinate and two local Lorentz indices, we follow
the supergravity convention of putting the manifestly-antisymmetric
coordinate-index pair to the left, $R_{\mu\nu}{}^{ab}=
  2\del_{[\mu}\omega_{\nu]}^{ab} +\cdots$.  Since in the former case all
the indices on the Riemann tensor are greek, whereas in the latter case
there are two greek and two latin indices, there shoud be no confusion.}
\begin{equation}
\widetilde R^\alpha{}_{\beta\mu\nu} =
\del_\mu \widetilde \Gamma^\alpha_{\nu\beta} -
\del_\nu \widetilde \Gamma^\alpha_{\mu\beta} +
\widetilde\Gamma^\alpha_{\mu\gamma} \widetilde\Gamma^\gamma_{\nu\beta}-
\widetilde\Gamma^\alpha_{\nu\gamma} \widetilde\Gamma^\gamma_{\mu\beta}\,.
\end{equation}
Thus we have
\begin{equation}
\widetilde R^{\alpha\beta}{}_{\mu\nu}= R^{\alpha\beta}{}_{\mu\nu} +
 \nabla_{[\mu} H_{\nu]}{}^{\alpha\beta} - \ft12 H^\alpha{}_{\lambda[\mu}
   H^{\beta\lambda}{}_{\nu]}\,.\label{riemanntilde}
\end{equation}
The Chern-Simons term ${\cal L}_{CS}$ can be written as the 6-form
\begin{equation}
{\cal L}_{CS}= -2 \widetilde\Theta^{\alpha\beta}\wedge \widetilde
\Theta_{\alpha\beta}\wedge B\,,
\end{equation}
where $\widetilde\Theta^{\alpha\beta}=
\ft12\widetilde R^{\alpha\beta}{}_{\mu\nu} \,dx^\mu\wedge dx^\nu$ is the
curvature 2-form with torsion.  Up to a total derivative, it may also be
written as
\begin{equation}
{\cal L}_{CS} = 2 \, \widetilde I_3\wedge H\,,
\end{equation}
where
\begin{equation}
d\widetilde I_3 = \tr(\widetilde\Theta\wedge\widetilde\Theta)=
  -\widetilde\Theta^{\alpha\beta}\wedge \widetilde
\Theta_{\alpha\beta}\,.
\end{equation}
The Chern-Simons form $\widetilde I_3$ is given by
\begin{eqnarray}
\widetilde I_3 &=& (\widetilde \Gamma^\alpha{}_{\mu\beta} \del_\nu
\widetilde\Gamma^\beta{}_{\rho\alpha} + \ft23
\widetilde \Gamma^\alpha{}_{\mu\beta} \widetilde \Gamma^\beta{}_{\nu\gamma}
\widetilde \Gamma^\gamma{}_{\rho\alpha})\,d x^\mu\wedge dx^\nu\wedge dx^\rho
\nn\ww2
&=&2 \widetilde{\cal L}_{LCS}\,  d^3 x\ ,
\end{eqnarray}
where we have used $d x^\mu\wedge dx^\nu\wedge dx^\rho=
 -\epsilon^{\mu\nu\rho} d^3 x$ and defined
the Lorentz-Chern-Simons Lagrangian
\begin{equation}\label{LCS}
\widetilde {\cal L}_{LCS} = - \ft12 \varepsilon^{\mu\nu\rho}
\left( \widetilde \Gamma^\alpha{}_{\mu\beta} \del_\nu
\widetilde\Gamma^\beta{}_{\rho\alpha} + \ft23
\widetilde \Gamma^\alpha{}_{\mu\beta} \widetilde \Gamma^\beta{}_{\nu\gamma}
\widetilde \Gamma^\gamma{}_{\rho\alpha} \right)\ .
\end{equation}
Under a variation of the connection, we have
\begin{equation}
\delta \widetilde I_3 = \delta\widetilde\Gamma^\alpha{}_{\mu\beta}
\widetilde R^\beta{}_{\alpha\nu\rho} \, dx^\mu\wedge dx^\nu\wedge dx^\rho
   + d\delta\nu\ ,
\end{equation}
where
\begin{equation}
\delta\nu = \delta\widetilde\Gamma^\alpha{}_{\mu\beta} \widetilde
\Gamma^\beta{}_{\nu\alpha} \, dx^\mu\wedge dx^\nu\ .
\end{equation}
Thus in terms of components, we have
\begin{equation}
\delta\widetilde I_{\mu\nu\rho} =
6 \widetilde R^\beta{}_{\alpha[\mu\nu}
\delta\widetilde\Gamma^\alpha{}_{\rho]\beta}
+ 3\del_{[\mu} \delta\nu_{\nu\rho]}\ .
\end{equation}

   We find that the equations of motion are given by
\begin{eqnarray}
&& e^{-2\phi}\, (R_{\mu\nu} + 2\nabla_\mu\nabla_\nu\phi -
 \ft14 H_{\mu\rho\sigma}
H_\nu{}^{\rho\sigma}) + \ft14 E_{\mu\nu} =0\ ,
\ww6
&&\nabla_\mu(e^{-2\phi} H^{\mu\nu\rho})
-12\alpha \nabla_\mu \widetilde\nabla_\sigma
 \widetilde R^{[\mu\nu\rho]\sigma}  +
 6\alpha \nabla_\mu \Big(\widetilde R^{[\mu\nu}{}_{\sigma\lambda}\,
  H^{\rho]\sigma\lambda}\Big)\nn\\
&&-\ft12\beta
\widetilde R^\alpha{}_{\beta\mu\sigma}\,
\widetilde R^\beta{}_{\alpha\lambda\tau}\,
\varepsilon^{\mu\sigma\lambda\tau\nu\rho}
 -6\beta \nabla_\mu \Big(\widetilde R^{[\mu\nu}{}_{\sigma\lambda}\,
  {^*\!H}^{\rho]\sigma\lambda}\Big)=0\,,\\
&&\nn\\
&&R - 4(\nabla\phi)^2 + 4\, \square\phi -\ft1{12} H^2 =0\ ,
\end{eqnarray}
where
\begin{eqnarray}
E_{\mu\nu} &=& 2\alpha \widetilde R^{\alpha\beta}{}_{\mu\rho}\,
  \widetilde R_{\alpha\beta\nu}{}^\rho
 -\ft12 \alpha \widetilde R^{\alpha\beta\rho\sigma}
\widetilde R_{\alpha\beta\rho\sigma}\, g_{\mu\nu}
-4\alpha \nabla_\alpha\widetilde\nabla^\lambda
  \widetilde R^\alpha{}_{(\mu\nu)\lambda} \nn\\
&&- 2\alpha
 \Big(\widetilde\nabla_\lambda
 \widetilde R^\alpha{}_{(\mu}{}^{\lambda\sigma}\Big)\,
  H_{\nu)\sigma\alpha}
+2\alpha \nabla_\beta\Big(\widetilde R^\beta{}_{(\mu}{}^{\alpha\gamma}
   H_{\nu) \alpha\gamma}\Big)
  -\alpha \widetilde R^\beta{}_{(\mu}{}^{\rho\lambda}
H_{\nu)\gamma\beta} H^\gamma{}_{\rho\lambda}\nn\\
&&
-2\beta \nabla_\beta\Big(\widetilde R^\beta{}_{(\mu}{}^{\alpha\gamma}
  \, {^*\!H}_{\nu) \alpha\gamma}\Big)
  +\beta \widetilde R^\beta{}_{(\mu}{}^{\rho\lambda}
H_{\nu)\gamma\beta} \, {^*\!H}^\gamma{}_{\rho\lambda}
\end{eqnarray}
and ${^*\!H}^{\mu\nu\rho}\equiv \ft16 \varepsilon^{\mu\nu\rho\sigma\lambda\tau}
H_{\sigma\lambda\tau}$.  (Note that the occurrence of some covariant
derivatives with torsion, and others without torsion, is intended, and is not
a misprint.)


\section{3-Sphere Reduction to Three Dimensions}


\subsection{The full bosonic action in three dimensions}


We now consider the 3-sphere reduction, with the ansatz given by
\begin{equation}\label{redans}
ds_6^2 = ds_3^2 + d\Sigma_3^2\ ,\qquad
H_3=2S\,\epsilon_\3 + 2m \Sigma_\3\ ,
\end{equation}
where $m$ is a constant, and
$d\Sigma_3^2$ is the metric of the round $S^3$ with $R_{ij}=
2m g_{ij}$.  Substituting the ansatz into the six-dimensional equations
of motion (with the parameter $\sigma$ introduced by sending
$\phi\longrightarrow \phi -\ft12\log\sigma$, as discussed previously),
we obtain equations of motion for the
three-dimensional fields:
\begin{eqnarray}
0&=& \alpha \, \square S + \sigma e^{-2\phi}S +m - \ft12 (\beta m- \alpha S)
(R+ 6S^2) \ ,
\label{S}
\ww2
0&=& \sigma \left( 4\, \square\phi -4(\del\phi)^2 +  R + 2S^2 +4m^2\right)\ ,
\label{phi}
\ww2
0&=& \sigma e^{-2\phi} \left[R_{\mu\nu} +2\nabla_\mu\nabla_\nu \phi
\right] - 2m S g_{\mu\nu}
\nn\ww2
&&
+\alpha\Bigg[ \square R_{\mu\nu}
-\ft12 \nabla_\mu\nabla_\nu R -4 R_\mu{}^\lambda R_{\lambda\nu}
+\ft52 RR_{\mu\nu} + \ft32 g_{\mu\nu}\left( R_{\rho\sigma}R^{\rho\sigma}
-\ft7{12}
R^2\right)
\nn\ww2
&&
\qquad\quad -\ft32 S^4 g_{\mu\nu} +G_{\mu\nu} S^2 -
\left(\nabla_\mu\nabla_\nu-g_{\mu\nu}\, \square\right) S^2
  -2\del_\mu S\, \del_\nu S +  (\del S)^2\, g_{\mu\nu}\Bigg]
\nn\ww2
&& +2\beta m \left[ S^3 g_{\mu\nu}-G_{\mu\nu} S +
\left(\nabla_\mu\nabla_\nu-g_{\mu\nu}\, \square\right) S -
    C_{\mu\nu}\right]\ ,
\label{Einstein}
\end{eqnarray}
where $G_{\mu\nu}= R_{\mu\nu} -\ft12 R g_{\mu\nu}$, and
$C_{\mu\nu}$ is the Cotton tensor defined by
\begin{equation}
C_{\mu\nu}= \varepsilon_\mu{}^{\rho\sigma} \nabla_\rho
\left(R_{\sigma\nu}-\ft14 g_{\sigma\nu} R\right)\ ,
\end{equation}
and the $\phi$ field equation has been used to simplify the
Einstein equation.
Note from (\ref{S}) that although $S$ is an auxiliary field in
the lowest-order theory (where $\alpha=0$ and $\beta=0$), it becomes dynamical
when $\alpha\ne0$.

It is useful to note, when performing the dimensional reduction, that
the connection with torsion in the 3-sphere directions is flat, and so
the curvature $\widetilde R^i{}_{jk\ell}$ on $S^3$ vanishes.

We find that the three-dimensional equations can be derived
from the Lagrangian
\begin{eqnarray}\label{d3lag}
\cL &=& \sigma e^{-2\phi} \left[R + 4 (\partial\phi)^2 +
4m^2  + 2S^2\right] + 4 m S - 2\beta m\left(R S +
2 S^3 - e^{-1}\, {\cal L}_{LCS}^{\rm Bos} \right)\cr
&&+\ft14 \alpha \left[ 4R_{\mu\nu}R^{\mu\nu} - R^2 -
   8 (\partial S)^2 + 12 S^4 + 4 RS^2\right]\ .
\end{eqnarray}
For later purposes, we shall write this as
\begin{equation}
{\cal L} = \sigma {\cal L}^{\rm Bos}_{EH}
+4m{\cal L}^{\rm Bos}_C  + \ft14 \alpha {\cal L}^{\rm Bos}_{Riem^2}
+2\beta m \left( {\cal L}^{\rm Bos}_{LCS}- {\cal L}^{\rm Bos}_{S^3}\right) \ ,
\label{mL}
\end{equation}
where
\begin{eqnarray}
\cL^{\rm Bos}_{EH} &=& e^{-2\phi} \left[R +  2S^2 + 4 (\partial\phi)^2 +
4m^2 \right]\ ,
\label{a1}\ww2
\cL^{\rm Bos}_C &=& S \ ,
\label{a2}\ww2
\cL^{\rm Bos}_{Riem^2}
&=& 4R_{\mu\nu}R^{\mu\nu} - R^2 - 8(\partial S)^2 + 12 S^4 + 4 RS^2\ ,
\label{a3}\ww2
e^{-1} {\cal L}^{\rm Bos}_{LCS} &=& \ft14 \varepsilon^{\mu\nu\rho} \left(
R_{\mu\nu}{}^{ab}\omega_{\rho ab} +\ft23 \omega_\mu{}^{a}{}_{b}
\, \omega_{\nu}{}^{b}{}_c \, \omega_{\rho}{}^{c}{}_{a}\right)\ ,
\label{a4}\ww2
\cL^{\rm Bos}_{S^3} &=&RS +2 S^3 \ ,
\label{a5}
\end{eqnarray}
where the spin connections and curvatures are both fermionic and
bosonic torsion-free.
(We write $\omega$ for $\omega(e)$ for simplicity in notation.)
It is worth remarking that if we had simply substituted the
ansatz (\ref{redans}) into the six-dimensional Lagrangian we would have
obtained the three-dimensional Lagrangian (\ref{d3lag}) but without
the $4mS$ term, and of course it would therefore not have given rise to
the correct three-dimensional equations of motion.  It is well known
that substituting a field-strength ansatz such as in (\ref{redans}) into
a higher-dimensional Lagrangian typically fails to give the correct
lower-dimensional Lagrangian.  It is interesting that the correct
Lagrangian is obtained, even when $H_{\mu\nu\rho}$ enters in higher-order
terms too, simply by adding the term $4mS$.

The higher-order terms in the Lagrangian,
proportional to $\alpha$ and $\beta$,
can simply be written as
\begin{equation}
\ft14 \alpha \widetilde R^{\mu\nu\rho\sigma} \widetilde R_{\mu\nu\rho\sigma}
   -2\beta m\, \widetilde{\cal L}^{\rm Bos}_{LCS}\ ,\label{2terms}
\end{equation}
where the tildes indicate, as usual the curvatures and connections
are those involving the bosonic torsion, as in (\ref{cminus}) with
$H_{\mu\nu\rho}=2 S \varepsilon_{\mu\nu\rho}$.  However, $S$ cannot simply
be absorbed as a torsion in the full theory, as can be seen even in
the leading-order bosonic terms $\cL^{\rm Bos}_{EH}$ and $\cL^{\rm Bos}_C$.
Furthermore, as we shall discuss below, the supersymmetric
completions of the bosonic terms in (\ref{2terms}) involve $S$-dependent
terms that cannot be absorbed as a torsion.

The three-dimensional  model we have obtained is an intriguing
mix of the off-shell supergravity multiplet and an on-shell
dilatonic scalar multiplet.  The combination of invariants in the
supergravity multiplet
is special case of the more general massive supergravity obtained in
\cite{behoroseto}.  However, our theory should not be viewed as trivial
generalization of the more general massive supergravity by adding a
matter coupling.  Truncating out the scalar multiplet in our theory
will not lead to the more general massive supergravity, but rather to
the trivial
Einstein-Hilbert term with a cosmological constant.  This can be seen
from the three-dimensional supersymmetry transformation rules,
which will be given in (\ref{susy3}) below.  Truncating
out $(\phi, \psi)$ requires us to take $S=-m$, and it follows from
(\ref{phi}) that the Ricci scalar becomes a constant.  Thus the matter
coupling in our model is more closely related to the supergravity
multiplet than a typical matter multiplet and the scalar multiplet
should be viewed as an integral part of the theory.

      The scalar $\phi$
in the three-dimensional multiplet has its origin in a mixing
of the six-dimensional dilaton and the breathing mode of the reduction ansatz.
Turning off the higher-order
derivative terms, the relevant Lagrangian is given by (we set $\sigma=1$
in the remainder of this subsection for simplicity)
\begin{equation}
\cL_{3} =
e^{-2\phi} (R +  2S^2 + 4 (\partial\phi)^2 + 4m^2) + 4\xi\,m S\ .
\end{equation}
Here, we have added a parameter $\xi$, which takes the value 0 or 1,
since $S$ is an independent invariant.  Integrating out the
auxiliary field $S$,
we have
\begin{equation}
\cL_3=e^{-2\phi} (R + 4(\partial\phi)^2 + 4m^2 - 2\xi m^2 e^{4\phi})\,.
\end{equation}
To see how this scalar $\phi$ arises as a mixing of the
six-dimensional dilaton and the breathing mode, let us examine the
six-dimensional Lagrangian in the Einstein frame,
\begin{equation}
{\cal L}_6=\ft12 \sqrt{-\hat g} \Big(\hat R -
\ft12 (\partial \hat \phi)^2
-\ft1{12} e^{-\sqrt2\hat\phi}\hat H^2_\3\Big)\,.
\end{equation}
The reduction ansatz including the breathing mode is given by
\begin{equation}
d\hat s_6^2 =e^{2 a \varphi} ds_3^2 + e^{2 b \varphi}
d\Sigma_3^2\ ,\qquad
\hat H_\3=2m(\epsilon_\3 + \xi\,\Sigma_\3)\ .
\end{equation}
where $a^2=\ft38 $ and $b=-\ft13 a$. Thus we have
\begin{eqnarray}
{\cal L}_3 &=& \sqrt{-g} \Big(R - \ft12(\partial \hat \phi)^2 -
\ft12(\partial\varphi)^2 - V\Big)\ ,\cr
V&=& 2m^2 (\xi\,e^{\sqrt2\hat \phi} + e^{-\sqrt2\hat\phi})
e^{4a\varphi} - 6m^2 e^{\ft83 a\varphi}\ ,
\end{eqnarray}
It turns out that we can make the consistent truncation,
\begin{equation}
\hat\phi = \ft12 \phi\,,\qquad \varphi=\sqrt2\, a \phi\,,
\end{equation}
so that the resulting Lagrangian is given by
\begin{eqnarray}
{\cal L} &=& \sqrt{-g} \Big (R - \ft12 (\partial\phi)^2 -
V\Big)\,,\cr
V &=& -2m^2 (2e^{\sqrt2\phi} - \xi\,e^{2\sqrt2\phi})\,.
\end{eqnarray}
The potential $V$ can be expressed in terms of a superpotential as
\begin{equation}
V= \Big(\fft{dW}{d\phi}\Big)^2 - W^2\,,\qquad
W=\sqrt2\,m(2e^{\ft1{\sqrt2} \phi} - \xi\,e^{\sqrt2\,\phi})\,.
\end{equation}
The reduction ansatz now becomes
\begin{equation}
ds_6^2 = e^{-\ft1{\sqrt8}\phi} \Big(e^{\ft1{\sqrt2}\phi} ds_3^2 +
d\Sigma_3^2\Big) = e^{-\ft1{\sqrt8}\phi}\Big(ds_{\rm str}^2 +
d\Sigma_3^2\Big)\ .
\end{equation}
It is clear that the vacuum solution AdS$_3\times S^3$ with $\xi=1$ is the decoupling limit of the self-dual string.  For $\xi=0$, the metric of the
vacuum solution is a domain wall, which is the decoupling limit of the electric string.


\subsection{The three-dimensional supersymmetry transformations}


Upon reduction to three dimensions, the supersymmetry parameter $\eps^i$, which is a
symplectic Majorana spinor, turns into a spinor $\eps^{iA}$ where
the $SO(2,1)$ spinor index as well as the spinor index on which the $\Sigma$ matrices
act are suppressed, while the $SO(3)$ spinor index
$A$ is exhibited. This spinor has $8$ real components, and therefore
it is associated with $\cN=4$ supersymmetry in three dimensions.
We shall truncate the theory to $\cN=1$ by setting
\begin{equation}
\epsilon^{iA} =\fft1{\sqrt2}\, \epsilon\, \Omega^{iA}\ .
\label{ka}
\end{equation}
The six-dimensional chirality condition now translates into
$\tau_3 \epsilon=\epsilon$, and the six-dimensional symplectic
Majorana condition becomes $\epsilon^*=-\im\, \epsilon$.

In the reduction to three dimensions, we shall let $\mu \to (\mu,\mu')$
and $a\to (a,a')$
where the primes are used in labeling the internal coordinate world
and Lorentz vectors.
In the bosonic sector we truncate as in \eq{tr1} and use ansatz
\eq{redans}, while in
the fermionic sector we set
\begin{equation}\label{tr2}
\psi_\mu^{iA} = \fft1{\sqrt2}\, \psi_\mu \Omega^{iA}\ ,\qquad
\psi^{iA}= \fft{\im}{\sqrt2}\, \Sigma_2 \psi \,\Omega^{iA}\ ,
\qquad \psi_{\mu'}^{iA}=0\ .
\end{equation}
As a consequence, we are left with the three-dimensional fields
$(e_\mu^a, \psi_\mu, S)$ and $(\psi, \phi)$. We find their supersymmetry
transformations to be

\begin{eqnarray}
\delta e_\mu^a &=& \ft12 \beps\gamma^a \psi_\mu\ ,
\nn\ww2
\delta \psi_\mu &=& D_\mu(\omega_-) \eps= D_\mu\ep +\ft12\gamma_\mu \ep\, S\ ,
\nn\ww2
\delta S &=& \ft{1}{8} \beps \gamma^{\mu\nu} \psi_{\mu\nu}(\omega_-)=
     \ft18\bar\ep\gamma^{\mu\nu} \psi_{\mu\nu} -
  \ft14 \bar\ep \gamma^\mu\psi_\mu\, S\ ,
\nn\ww2
\delta \psi &=& \ft14\,
e^{5\phi/4} \left( \gamma^\mu\partial_\mu \phi + S +m \right)\eps\ ,
\nn\ww2
\delta\phi &=& e^{-5\phi/4} \beps \psi\ ,
\label{susy3}
\end{eqnarray}
where $\psi_{\mu\nu}(\omega_-)=2D_{[\mu}(\omega_-) \psi_{\nu]}$ and
\begin{equation}
\omega_{\mu\pm}{}^{ab} = \omega_\mu{}^{ab} \pm \varepsilon_\mu{}^{ab} S\ ,
\label{torsion}
\end{equation}
Using \eq{redans}, we find that $\omega_{\mu'-}{}^{a'b'}=0$
which simplifies the reduction formulae considerably.
For example, $\psi_{\mu\nu'}=0$ and $\psi_{\mu'\nu'}=0$.

Our results for the transformation rules for $(e_\mu^a,\psi_\mu,S)$
agree precisely with the known off-shell $\cN=1$ supergravity
multiplet transformations in three dimensions. The field $S$ admits a
torsion interpretation \cite{anbedrhoseto}.

In section 3.1, we explained why the scalar field $S$ (and its fermionic
partner) cannot be truncated away in presence of the higher derivative
couplings, by considering the field equations.
The supersymmetry transformation rules above provide another simple
explanation of this phenomenon as follows. Setting $\phi=0$ implies that
$S=-m$. But then the supersymmetry variation of $S$ implies the gravitino
field equation without higher derivative terms. Hence, the higher
derivative couplings must be absent altogether if we are to be able to
truncate out the scalar multiplet.


\subsection{The supersymmetric completion of ${\cal L}_{EH}^{\rm Bos}$
and ${\cal L}_C^{\rm Bos}$}


The supersymmetric completion of ${\cal L}_{EH}^{\rm Bos}$ and
${\cal L}_C^{\rm Bos}$
can be obtained by performing
the 3-sphere reduction of the off-shell Poincar\'e sector of the
six-dimensional theory. Since the fermionic sector of this theory has not
been provided until now, we construct the supersymmetric completion directly
in three dimensions, by starting from the bosonic sector and
supersymmetry transformation rules we obtained
from  the 3-sphere reduction. We find, up to quartic fermion terms,
\begin{eqnarray}
\cL_{EH} &=&
 e^{-2\phi} \left[R +  2S^2 + 4 (\partial\phi)^2 +
4m^2 \right]
\nn\ww2
&& + e^{-2\phi}\, \Big[- \bar\psi_\mu R^\mu
+ 2\bar\psi^\mu\gamma^\nu \psi_\nu\, \del_\mu\phi +
 m\bar\psi_\mu\gamma^{\mu\nu} \psi_\nu\Big]\nn\\
&&-8\, e^{-\ft{13}{4}\phi}\, \Big[\bar\psi \gamma_\mu R^\mu
+ \bar\psi\gamma^\mu\gamma^\nu \psi_\mu \del_\nu\phi
 + m\bar\psi_\mu\gamma^\mu\psi\Big]\nn\\
&& +8 e^{-\ft92\phi}\,
\Big[ \bar\psi\psi\, S + 2 \bar\psi\gamma^\mu D_\mu\psi -2m \bar\psi \psi\Big]\ ,
\label{EH}\ww4
e^{-1} {\cal L}_C &=& S +\ft18 {\bar\psi}_\mu
\gamma^{\mu\nu}\psi_\nu \  .
\end{eqnarray}
%


\subsection{The supersymmetric completion of ${\cal L}_{Riem^2}^{\rm Bos}$,
${\cal L}_{LCS}^{\rm Bos}$ and ${\cal L}_{S^3}^{\rm Bos}$}


   Using \eq{tr1},
\eq{redans} and \eq{tr2}, we find that the 3-sphere reduction of the
six-dimensional Lagrangian ${\cal L}_{Riem^2}$ given in
\eq{riem2} yields\footnote{The reduction of the
second term in  \eq{riem2} gives rise to
${\bar\psi}^{ab} {\slashed D}(\omega,\omega_+)\psi_{ab}$ with the covariant
derivative defined in \eq{sdef2} for an $Sp(1)$ singlet. We convert that
to ${\bar\psi}^{ab} {\slashed D}(\omega_-)\psi_{ab}$ with the covariant
derivative defined in \eq{covder1} by adding and subtracting the required
terms. Our result corrects that of \cite{anbedrhoseto} for the
${\rm Riemann}^2$ invariant, where
${\bar\psi}^{ab}{\slashed D}(\omega)\psi_{ab}$ is used, instead of
${\bar\psi}^{ab} {\slashed D}(\omega,\omega_+)\psi_{ab}$.}

\begin{eqnarray}\label{riem33}
\cL_{Riem^2}^{D=6} \ \xrightarrow{S^3}\ &&  \cL_{Riem^2}
+\cL_{LCS} + \cL_{S^3}
\nn\ww2
&=& \bigg[R_{\mu\nu ab}(\widehat \omega_+)
R^{\mu\nu ab}(\widehat \omega_+) +2{\bar\psi}^{ab}(\omega_-)
\slashed D(\omega_-) \psi_{ab}(\omega_-)
\nn\ww2
&& -R^{\mu\nu ab}(\omega_-) {\bar\psi}_\lambda \Gamma^{ab}
\Gamma^\lambda\psi_{\mu\nu}(\omega_-)
+2S {\bar\psi}_{\mu\nu}(\omega_-)\gamma_\rho\gamma^\mu \psi^{\nu\rho}(\omega_-)\bigg]
\nn\ww2\
&&-8m\left[ \ft14 \varepsilon^{\mu\nu\rho} \left(
R_{\mu\nu}{}^{ab}(\hat\omega){\hat\omega}_{\rho ab} +\ft23 {\hat\omega}_\mu{}^{a}{}_{b}
{\hat\omega}_{\nu}{}^{b}{}_c {\hat\omega}_{\rho}{}^{c}{}_{a}\right)
+\ft12 {\bar R}^\mu(\omega)\gamma_\nu\gamma_\mu R^\nu(\omega)\right]
\nn\ww2
&& +8m\bigg[R(\omega)S +2 S^3
-\ft12 {\bar R}^\mu(\omega) \gamma_\mu \gamma_\nu R^\nu(\omega)
- \bar\psi_\mu\gamma^\mu\psi^\nu\, \del_\nu S
\nn\ww2
&& +\ft12 {\bar\psi}_\mu \gamma^{\mu\nu} R_\nu (\omega) S
-\ft12 \bar\psi^\mu\psi_\mu S^2\bigg]\,,
\end{eqnarray}
modulo the quartic fermion terms in the sense described earlier. Furthermore
\begin{eqnarray}\label{omegahat3}
D_\mu(\omega_-)\psi_{ab} &=&\left( \partial_\mu
  +\ft14\omega_{\mu-}{}^{cd}\Gamma_{cd}\right)\psi_{ab}
  -2\omega_{\mu-[a}{}^c \psi_{b])c}\ ,
\label{covder1}\ww2
{\widehat\omega}_{\mu+}{}^{ab} &=&
\omega_\mu{}^{ab} +\ft12{\bar\psi}_\mu\gamma^{[a}\psi^{b]}
+\ft14 {\bar\psi}^a\gamma_\mu\psi^b + \varepsilon_\mu{}^{ab} S\ ,
\ww2
{\widehat\omega}_{\mu}{}^{ab} &=&
\omega_\mu{}^{ab} +\ft12{\bar\psi}_\mu\gamma^{[a}\psi^{b]}
 +\ft14 {\bar\psi}^a\gamma_\mu\psi^b \ ,
\ww2
\omega_{\mu-}{}^{ab} &=& \omega_\mu{}^{ab} - \varepsilon_\mu{}^{ab} S\,.
\end{eqnarray}
We have grouped the terms in (\ref{riem33}) as a sum of three terms,
each enclosed in square brackets.  Each bracketed set is
separately invariant under three-dimensional ${\cal N}=1$ supersymmetry.
The terms in the first bracket furnish
a supersymmetrization of the Riemann tensor
squared term (with bosonic torsion).
The second set of bracketed terms agrees with the topologically massive
supergravity action ${\cal L}_{LCS}$ that has been known for some time
\cite{deskay}. The third bracket provides a superextension of the combination
$RS+2S^3$.  Although the existence of such a super-invariant
had been noted in \cite{Bergshoeff:1987rb}, its explicit form has not
previously been given.

   Up to quartic fermion terms, the Lagrangians for the three
super-invariants can be written more explicitly as
\begin{eqnarray}
e^{-1}{\cal L}_{Riem^2} &=& 4R_{\mu\nu}R^{\mu\nu}- R^2
-8\partial^\mu S\partial_\mu S + 4RS^2 +12 S^4
\nn\ww2
&&  +4{\bar\psi}^\mu\gamma^\nu\psi^\rho \nabla_\rho R_{\mu\nu}
+ \psi^\mu\gamma^{\nu\rho}\psi_\rho \left( R_{\mu\nu}-4\nabla_\mu \partial_\nu S \right)
-2{\bar\psi}^\mu\gamma^\nu\psi_\nu \partial_\mu S^2
\nn\ww2
&&  +\ft18 {\bar\psi}_\mu\gamma^{\mu\nu}\psi_\nu \left(3RS + 8\square S +16 S^3\right)
+2{\bar\psi}_{ab}(\omega_-) \slashed D(\omega_-) \psi_{ab}(\omega_-)
\nn\ww2
&&  -R^{\mu\nu ab}(\omega_-) {\bar\psi}_\lambda \Gamma^{ab}
\Gamma^\lambda\psi_{\mu\nu}(\omega_-)
+2S {\bar\psi}_{\mu\nu}(\omega_-)\gamma_\rho\gamma^\mu \psi^{\nu\rho}(\omega_-)\ ,
\label{R22}\ww4
e^{-1} {\cal L}_{LCS} &=&\ft14\varepsilon^{\mu\nu\rho} \left(
R_{\mu\nu}{}^{ab}\omega_{\rho ab} +\ft23 \omega_\mu{}^{a}{}_{b}
\omega_{\nu}{}^{b}{}_c \omega_{\rho}{}^{c}{}_{a}\right)
+\ft12{\bar R}^\mu \gamma_\nu\gamma_\mu R^\nu
\nn\ww2
&& +\ft12 \varepsilon^{\mu\nu\rho}
\left(R_{\rho\sigma}-\frac14 g_{\rho\sigma} R\right)
{\bar\psi}_\nu \gamma^\sigma\psi_\rho\ ,
\label{LCS2}
\ww4
e^{-1}{\cal L}_{S^3} &=&
RS +2 S^3
-\ft12 {\bar R}^\mu \gamma_\mu \gamma_\nu R^\nu
- \bar\psi_\mu\gamma^\mu\psi^\nu\, \del_\nu S
\nn\ww2
&& +\ft12 {\bar\psi}_\mu \gamma^{\mu\nu} R_\nu S
-\ft12 \bar\psi^\mu\psi_\mu S^2\,,
\end{eqnarray}
where all curvatures in which the arguments are not indicated are
understood to be torsion-free.


\section{Generalization of the Model in Three Dimensions and its Critical
Points}


The three dimensional model we have obtained through the 3-sphere
reduction from
six dimensions has two continuous parameters, namely the
cosmological constant $m^2$, and the coupling
constant $\alpha$ in front of the Riemann squared action, which are both
dimensionless if measured in units of $\kappa$ (which we set to unity, for
convenience).  We can also include the discrete parameter $\sigma$, the
coefficient of the Einstein-Hilbert term, taking the values 1, 0 or
$-1$.  Since the
three-dimensional $\cN=1$ supersymmetry is less restrictive than the
original six-dimensional
$\cN=(1,0)$ supersymmetry, we can generalise the three-dimensional
$\cN=1$ theory to include eight parameters, with the Lagrangian in the
bosonic sector given by
\begin{eqnarray}\label{genmod}
{\cal L} &=&\sigma  e^{-2\phi} \left[R +  2S^2 + 4 (\partial\phi)^2 +
4m^2 \right] +M{\cal L}^{\rm Bos}_C
\nn\ww2
&& + \ft14 \alpha {\cal L}^{\rm Bos}_{Riem^2}
+2\beta m \left( {\cal L}_{LCS}^{\rm Bos}-
a {\cal L}_{S^3}^{\rm Bos}\right)
+ b{\cal L}_{R^2}^{\rm Bos} + c{\cal L}^{\rm Bos}_{S^4}\ ,
\end{eqnarray}
where   ${\cal L}^{\rm Bos}_C$, ${\cal L}^{\rm Bos}_{Riem^2}$,
${\cal L}^{\rm Bos}_{LCS}$ and
${\cal L}^{\rm Bos}_{S^3}$ are as defined in \eq{a1},
\eq{a2}, \eq{a3}
and \eq{a4}, while the last two terms in the Lagrangian are given
by \cite{Bergshoeff:1986wc,Bergshoeff:1987rb}
\begin{eqnarray}\label{newLag}
\cL_{R^2}^{\rm Bos} &=& R^2 -16(\del S)^2 +12 RS^2 +36 S^4\ ,
\ww2
\cL_{S^4}^{\rm Bos} &=&  3RS^2 +10 S^4\ .
\end{eqnarray}
Note that we have set $\kappa^2=1$ and introduced the new positive or
negative
real parameters $M,a,b,c$.  Thus the count of
eight parameters comprises seven real dimensionless parameters (measured
in units of $\kappa$), and the discrete parameter
$\sigma=\pm 1,0$.\footnote{One could take the view that $\sigma$ should
not strictly speaking be thought of as a non-trivial parameter in the
theory, since, as noted in section 2,
any value of $\sigma$ can be obtained from $\sigma=1$ by
means of the field transformation $\phi\longrightarrow \phi -\ft12 \log\sigma$.
However, it is useful to include it explicitly in the Lagrangian since
the cases where $\sigma$ is negative and $\sigma=0$ have properties
that are physically distinct from the case when $\sigma$ is positive. Thus
perhaps the most useful viewpoint is that there are three distinct
seven-parameter
theories, corresponding to $\sigma=+1$, $\sigma=-1$ and $\sigma=0$.}
Compared to the seven-parameter model of \cite{Bergshoeff:1987rb} our extra
parameter is $m$.\footnote{The counting of seven parameters in
\cite{Bergshoeff:1987rb} also includes the discrete constant $\sigma$, and
again, one could perhaps most appropriately view the model as comprising
three distinct six-parameter theories.  Although
there is no $\phi$ field in the theory in \cite{Bergshoeff:1987rb}, $\sigma$
is again in a sense a ``redundant'' parameter, since it
can be introduced, starting from the case $\sigma=1$, by means of appropriate
scaling transformations of the fields and the other coupling constants.}

We note that
\begin{equation}\label{6dsusy}
D=6 \ {\rm supersymmetry}\ \ \ \ \Longrightarrow  \ \ \ \ M=4m\ ,
\quad \beta= \alpha\ ,\quad a=1\ ,\quad b=c=0\ .
\end{equation}
It should be emphasised that the case with $\beta=-\alpha$ can also
be lifted to six-dimensional supergravity, provided that the supersymmetry
transformation rules and spinor chiralities are modified appropriately.

Turning to the generalized massive supergravity model, the bosonic part of
the full Lagrangian takes the form
\begin{eqnarray}
\label{fullbosonic}
\cL &=& \sigma e^{-2\phi}\left( R+4(\partial\phi)^2 +2S^2+4 m^2\right) +MS
+\alpha R_{\mu\nu}R^{\mu\nu} +\ft14(4b-\alpha) R^2
\nn\ww2
&& -2 (\alpha+8b) (\partial S)^2 +(3\alpha+36 b +10c) S^4
+(\alpha+12b+3c) RS^2
\nn\ww2
&& -2\beta a (RS +2S^3) +2\beta m \cL_{LCS}\ .
\end{eqnarray}

For generic
values of these parameters, the fluctuations around the AdS$_3$ vacuum
solution is expected to describe
two helicity $|\nu|=2$ states and three scalars, the latter coming from the
trace of the metric, the auxiliary field $S$ and the dilatonic scalar
$\phi$. For special values of the coupling constants, however, some or
all of the helicity $|\nu|=2$ states may become singletonic in the sense
that they become confined to propagate on the boundary of AdS$_3$.
Additionally, it may be possible that
the trace of the metric can be gauged away by residual coordinate
transformations.  The massive supergravity model with bosonic sector
given in \eq{genmod}
differs from that studied recently in \cite{Bergshoeff:1987rb} owing to the
replacement of $R-2S^2$, which is the bosonic sector of
the simple supergravity, by the Lagrangian \eq{a1}. One of the consequences of
doing so is that we have one extra parameter in the full Lagrangian.

We expand the metric around the AdS$_3$ background as
$g_{\mu\nu}={\bar g}_{\mu\nu} +h_{\mu\nu}$, and impose the gauge condition
\begin{equation}\label{gc}
\nabla^\mu H_{\mu\nu} =0\ ,\quad \hbox{where}\quad
H_{\mu\nu} \equiv  h_{\mu\nu} - \ft{1}{3} \bar g_{\mu\nu} h\ ,
\qquad h \equiv h_{\mu\nu} {\bar g}^{\mu\nu}\ .
\end{equation}
We expand the scalar fields $S$ and $\phi$ around the supersymmetric  vacuum
solution $\bar S=-m$ and $\bar \phi=0$, and denote the fluctuation fields by
$s$ and $\phi$, respectively. The requirement that the $S$ field equation
be satisfied by the vacuum solution implies that
\begin{equation}\label{vac}
M=4m(\sigma +m^2 c)\ .
\end{equation}
We shall use this relation in subsequent calculations to eliminate $M$.
Setting $m=1$ from here on for simplicity, the resulting
field equations for the scalar fields $\phi$, $s$, and the trace of the
Einstein equation, respectively, take the form
\begin{eqnarray}
 6\sigma \, \square \phi - \sigma (\square -3) h-6\sigma s &=&0\ ,
\label{s1}\ww2
3\left(\alpha+8b\right) \square s +3\left(\sigma+3c+6\gamma\right) s
+\gamma\left(\square-3\right) h +6\sigma\phi &=&0\ ,
\label{s2}\ww2
\left(\square-3\right)Y &=&0\ ,
\label{s3}
\end{eqnarray}
where $\square$ is defined in the AdS$_3$ background, and
\begin{eqnarray}\label{Y}
Y &=& \left(\alpha+8b\right) \left(\square -3\right) h -\gamma h
+12\left(\rho s+\sigma \phi \right)\ ,
\nn\ww2
\gamma &\equiv & \sigma+3c-\alpha+2\beta a\ ,
\nn\ww2
\rho &\equiv&  12b +3c + \alpha + \beta a \ .
\end{eqnarray}
The $s$ field equation differs from \cite{Bergshoeff:1987rb} only in
presence of the scalar field $\phi$ {\emph and} the sign of $\sigma$.

   As commented upon earlier, the scalar $\phi$ couples to the supergravity
multiplet in a non-trivial way.  It follows that, unless $\sigma=0$,
there is no choice of parameters for which all three scalars can be eliminated
by using their equations of motion. Consequently, the discussion of the
unitarity
of the theory is more complicated than the discussion in \cite{behoroseto},
where, for a certain choice of parameters, $s$ and $h$ could be eliminated
allowing a unitarity condition to be obtained. If $\sigma=0$, the
partial results of  \cite{Bergshoeff:1987rb} on perturbative unitarity can be
used.
With the parameters required by $D=6$ supersymmetry, however,
one does not obtain a
model for which a conclusion about perturbative unitarity can be drawn, based
on the results of \cite{Bergshoeff:1987rb} alone.

There remains the traceless part of the Einstein equation, which can be
expressed as
\begin{equation}
\cD(1)\cD(-1)\cD(\eta_+) \cD(\eta_-) H_{\mu\nu}= \gamma^{-1}\, J_{\mu\nu} \ ,
\label{key1}
\end{equation}
where we have defined the differential operator\footnote{Note that by
definition $\cD(\eta) H_{\mu\nu}$ means $[\cD]_\mu{}^\rho H_{\rho\nu}$,
and that, as can easily be verified, this preserves the transversality
and tracelessness of $H_{\mu\nu}$.  Furthermore, the
operators commute on $H_{\mu\nu}$, in the sense that
$[\cD(\eta_1),\cD(\eta_2)]H_{\mu\nu}=0$ for any $\eta_1$ and $\eta_2$. Another
useful identity satisfied by these operators
is that $D(\eta_1) D(\eta_2) H_{\mu\nu}=
\eta_1\eta_2\, (\,\square +3)H_{\mu\nu} + \cD(\eta_1+\eta_2) H_{\mu\nu}$.}
\begin{equation}
\left[{\cal D}(\eta)\right]_\mu{}^\nu = \delta_\mu{}^\nu
+ \eta\, \varepsilon_\mu{}^{\alpha\nu}\nabla_\alpha\ ,
\label{firstorder}
\end{equation}
for a constant $\eta$, and
\begin{equation}\label{eta}
\eta_\pm = \gamma^{-1}
\left(- \beta\pm \sqrt{\beta^2-\gamma\alpha}\,\right)\ .
\end{equation}
Note that this result is independent of the parameter $b$ that occurs in
front of
${\cal L}_{R^2}$ in the total Lagrangian. We have assumed that
$\gamma\ne 0$. The source term is given by
\begin{equation}\label{J}
J_{\mu\nu}= -\ft13 \left(\nabla_\mu \nabla_\nu -\ft13 \gb_{\mu\nu} \,
\square\right) Y\ .
\end{equation}
These satisfy $\eta_+\eta_- = \alpha\gamma^{-1}$. The integrability condition $\nabla^\mu J_{\mu\nu}=0$ is satisfied by virtue
of the field equation \eq{s3}. Provided that $\gamma\ne 0$, the
source term $J_{\mu\nu}$ can be absorbed into the definition of $H_{\mu\nu}$ in such
a way that it maintains the traceless and transverse properties of $H_{\mu\nu}$.

   For appropriate boundary conditions, the vanishing of the left-hand side
of (\ref{key1}) implies that $H_{\mu\nu}$ is annihilated by one or another
of the
four commuting $\cD$ factors.  In general, the helicity $\nu$ and
lowest energy $E_0$
of an excitation satisfying $\cD(\eta) H_{\mu\nu}=0$ are given by
\cite{behoroseto}
\be
\nu= \fft{2\eta}{|\eta|}\,,\qquad E_0 = 1+\fft1{|\eta|}\,,
\ee
and the mode furnishes a unitary irreducible representation of
the AdS$_3$ group if $E_0\ge |\nu|$, which means $|\eta|\le 1$.
If $\eta=\pm1$, the mode
decouples in the bulk, and just describes an excitation in the boundary
theory.  Thus generically, when $|\eta_\pm|\ne 1$, there are two bulk
graviton modes with $|\nu|=1$ and two boundary modes.

The expression \eq{eta}, for $\gamma\ne0$, agrees with that found for
the seven-parameter action in
\cite{Bergshoeff:1987rb}. The additional
parameter that we have in our action and the coupling of the scalar multiplet
does not change this result, because upon expansion
around the vacuum solution we have employed, the linearization of
$\sqrt{-g}\left[ \sigma e^{-2\phi} (R+2S^2 +4m^2) +MS\right]$ gives the
same result as $\sqrt{-g}\left[\sigma (R-2S^2) +M S\right]$ in
\cite{Bergshoeff:1987rb}. As we saw earlier, however, the field equations
for the scalar fields do differ in the two cases.

   For $\gamma=0$, a straightforward limit of (\ref{key1}) gives
\be
\cD(1)\cD(-1)\cD(\eta) \varepsilon_\mu{}^{\alpha\beta}\nabla_\alpha
   H_{\beta_\nu} =-\fft1{2\beta}\, J_{\mu\nu}\,,
\ee
where $\eta=-\alpha/(2\beta)$.
The source term cannot be absorbed
into a redefinition of $H_{\mu\nu}$ in this case.  Upon acting with
$\varepsilon_\lambda{}^{\tau\mu}\,\nabla_\tau$, the source term
drops out, yielding \cite{behoroseto}
\be
\cD(1)\cD(-1) \cD(\eta)(\square+3) H_{\mu\nu}=0\,.
\ee
In addition to a single helicity 2 massive graviton, this equation
also describes a partially-massless graviton \cite{pm1,pm2,pm3}.

    The critical points where the massive graviton decouples arise
when either $|\eta_+|=1$ and/or $|\eta_-|=1$ and/or $\eta_+=\eta_-$,
with $\eta_\pm$ given in (\ref{eta}).  The criticality condition in
our eight-parameter model coincides with that in \cite{behoroseto}, where
an extensive list of critical points was given. For our three-parameter theory
that can be lifted to six dimensions, $a=1$, $b=0=c$ and the
critical points are given by
\bea
\sigma^2 =1:\qquad \hbox{Case 1:}&&
\beta=+\alpha:\qquad \sigma \alpha =-\ft1{4}\ ,\qquad \eta_+=1\ ,\nn\\
\hbox{Case 2:} &&
\beta=-\alpha:\qquad \sigma \alpha =+\ft1{4}\ ,\qquad\eta_+=\eta_-=1\ ,
\nn\ww4
\sigma=0: \qquad
\hbox{Case 3:} &&
{\rm Any}\ \alpha \ne \beta\ ,\qquad \eta_- = -1\ , \nn\\
\hbox{Case 4:} &&
{\rm Any}\ \alpha=\beta\ , \qquad \eta_+ = \eta_- = -1 \ .
\label{crits}
\eea
In Case 1 and Case 3, there are only single helicity $-2$ bulk states
with AdS energies $E_0=4$ and $E_0=1+|(2\beta-\alpha)/\alpha|$, respectively.
In Case 3, taking $\beta=-\alpha$
gives $E_0=4$ as well. In Case 2 and Case 4, there are no propagating
bulk gravitons at all.

Finally, we may evaluate the central charges for the right-handed and
left-handed Virasoro algebras of the boundary CFT, following the
procedure described in \cite{entropy1,entropy2,entropy3,behoroseto}, finding
\be
C_L= \fft{3}{2G} \, \Big( \sigma + 3 c + 2\beta(a + 1)\Big)\,,\qquad
C_R= \fft{3}{2G} \, \Big( \sigma + 3 c + 2\beta(a - 1)\Big)\,.
\ee
(We have restored Newton's constant in order to simplify comparison with
previous results.)
For our three-parameter theory coming from the reduction from six dimensions,
we have $a=1$ and $c=0$.   Thus for $\sigma^2=1$, we have $c_L=0$ if
$\sigma \beta =-\ft1{4}$, while $C_R=3\sigma/(2G)$.  This corresponds to the
critical points listed as Case 1 and Case 2 in (\ref{crits}).
If instead $\sigma=0$, then
$C_R=0$ for any $\beta$, and $C_L= 6\beta/G$. In this case,
$\beta$ can be chosen so that
$C_L$ has any desired value.  This may have interesting consequences for
the corresponding CFT. This case leads to the critical points listed as
Case 2 and Case 4 in (\ref{crits}).

   In summary, Case 2 can be viewed as the  higher-derivative
generalization of chiral gravity proposed in
\cite{lisost}, and Case 4 is an alternative higher-derivative version in
which the Hilbert-Einstein term is omitted, with the parameters chosen
in each case so that no massive gravity
modes arise. Moreover, this phenomenon occurs in the alternative theory
for any value of the parameters with $\alpha=\beta$.


\section{Dualisation and the Heterotic String}


   The six-dimensional supergravity whose bosonic Lagrangian is
given by (\ref{sezginlag}) is closely related to the dimensional
reduction of the effective theory of the heterotic string. To be
more precise, we may consider the ten-dimensional supergravity
constructed in \cite{bergsderoo}, where the supersymmetrisation
of the anomaly-canceling $\tr( R\wedge R)$ term in the
Bianchi identity for the 3-form $H_\3$ was studied.  The goal
in \cite{bergsderoo} was to consider only those terms that
are necessary in order to obtain a Lagrangian that remains supersymmetric
when the the Bianchi identity $dF_\3= \tr(F\wedge F)$ is modified to
$dF_\3= \tr(F\wedge F)- \alpha'\, \tr( R\wedge R)$.  It was shown
that this requires introducing higher-order curvature terms
in the Lagrangian, starting with $\ft14 \alpha e^{-2\phi} R^{\mu\nu\rho\sigma}\,
R_{\mu\nu\rho\sigma}$, and that furthermore the curvature in these
terms is built from the connection $\widetilde \Gamma^\mu{}_{\nu\rho} =
\Gamma^\mu{}_{\nu\rho} -\ft12 F^\mu{}_{\nu\rho}$ with bosonic torsion.
Supersymmetry requires that the Lagrangian with the anomaly-canceling
$\tr( R\wedge R)$ term have corrections to arbitrarily high order in
$\alpha'$ (and hence arbitrarily high powers of the curvature).  In
\cite{bergsderoo}, these corrections were studied up to and including
order ${\alpha'}^2$.

   If the theory of \cite{bergsderoo} is reduced on $T^4$ to six
dimensions (setting the Yang-Mills fields to zero, and consistently
truncating to keep only the relevant fields),
it can be compared with the theory of
\cite{Bergshoeff:1985mz,Bergshoeff:1986wc}, given in
(\ref{sezginlag}).  The reduced heterotic effective action gives
\begin{equation}
{\cal L}_6 =\sqrt{-g}\, e^{-2\phi}\,
 \Big[
 R + 4 (\del\phi)^2 - \ft1{12} F^{\mu\nu\rho} F_{\mu\nu\rho}
 +\ft14 \alpha
\widehat R^{\mu\nu\rho\sigma} \widehat R_{\mu\nu\rho\sigma}\Big]
 + {\cal O}(\alpha^2)\,,\label{hetlag}
\end{equation}
where here the curvature in the terms at order $\alpha$ is calculated
using the connection $\widehat \Gamma^\mu{}_{\nu\rho} =
\Gamma^\mu{}_{\nu\rho} -\ft12 F^\mu{}_{\nu\rho}$, and $F_\3$ is given by
\begin{equation}
F_\3 = dA_\2 - \ft12\alpha\, \widehat I_3\,.
\end{equation}

   If we neglect for a moment the torsion contributions to the higher-order
curvature terms in the two six-dimensional theories, it is easy to see
that (\ref{sezginlag}) and (\ref{hetlag}) are related by dualisation,
with the dilaton and metric of the theory (\ref{sezginlag}) transformed
according to
\begin{equation}
\phi\longrightarrow -\phi\,,\qquad g_{\mu\nu}\longrightarrow e^{-2\phi} \,
  g_{\mu\nu}\label{duality1}\,.
\end{equation}
(All fields on the right-hand sides are the transformed fields.)
Then we define (using the dualised dilaton and metric)
\begin{equation}
F_{\mu\nu\rho} = \ft16 \varepsilon_{\mu\nu\rho\alpha\beta\gamma}\, e^{2\phi}\,
                 H^{\alpha\beta\gamma}\,.\label{duality2}
\end{equation}
Note that under this dualisation, the Chern-Simons term ${\cal L}_{CS}$ in
(\ref{sezginlag}) gives rise to the anomaly-canceling $\tr( R\wedge R)$
term in the theory described by (\ref{hetlag}).

   If we include the torsion contributions, the duality between
(\ref{sezginlag}) and (\ref{hetlag}) is harder to see, but we conjecture
that it does still exist, and it implemented by exactly the same
transformations (\ref{duality1}) and (\ref{duality2}).  It should be
emphasised that in particular, this conjectured duality relates the
exact, closed-form, theory given by (\ref{sezginlag}) (which has no
curvature corrections beyond  ${\cal O}(\alpha)$) to the theory described
by (\ref{hetlag}) with its curvature corrections to all orders in $\alpha$.

   The strongest reason for believing this duality conjecture is that
the theory described by (\ref{sezginlag}) (together with its fermionic
terms as given in \cite{Bergshoeff:1985mz,Bergshoeff:1986wc}) is
exactly supersymmetric, with
${\cal N}=(1,0)$ supersymmetry.  On the other hand, the theory described
by (\ref{hetlag}) (together with its fermionic terms) is the dimensional
reduction of the supersymmetrisation of the anomaly-canceling
$\tr( R\wedge R)$ in ten dimensions, and this theory (after the reduction and
truncation we
have performed) also has ${\cal N}=(1,0)$ supersymmetry in six dimensions.
In each case, the supersymmetrisation procedure gave a unique result.
Since the dualisation of (\ref{sezginlag}) will certainly give rise
to {\it some} six-dimensional theory with an anomaly-canceling
$\tr( R\wedge R)$ term, the uniqueness of the constructions implies that
it can only be giving rise to the theory described by (\ref{hetlag}).

   A remarkable consequence of this duality is that the infinite set
of correction terms to all orders in $\alpha$ that are needed in order to
supersymmetrise the anomaly-canceling $\tr( R\wedge R)$ term in the
theory studied in \cite{bergsderoo}) can be deduced (modulo terms that
vanish in the $T^4$ reduction) simply by performing
the dualisation of the theory constructed in (\ref{sezginlag}), which is
exactly supersymmetric without the need for any corrections beyond the
order $\alpha$.

    Providing a complete proof of the duality would be quite involved, and
we shall not attempt it here.  Instead, we shall just focus on a
sub-calculation that is already non-trivial, and that elucidates a
seemingly puzzling aspect of the dualisation.  The puzzle is that
in both the formulation in (\ref{sezginlag}) and the formulation in
(\ref{hetlag}), the quadratic curvature terms are constructed from
a connection with bosonic torsion; $-\ft12 H^\mu{}_{\nu\rho}$ in the
case of (\ref{sezginlag}) and $-\ft12 F^\mu{}_{\nu\rho}$ in the case of
(\ref{hetlag}).  However, these fields are related by (\ref{duality2}), and
so one might have expected that if the (\ref{sezginlag}) theory had
torsion proportional to $H_{\mu\nu\rho}$ then the dual theory would
have torsion proportional to $\ep_{\mu\nu\rho\alpha\beta\gamma}
 F^{\alpha\beta\gamma}$ rather than $F_{\mu\nu\rho}$.  Here, we shall
look specifically at the contributions at linear order in $F_{\mu\nu\rho}$,
associated with the anomaly-canceling term $\tr( \widehat R\wedge
\widehat R)$, in the expression for the dualised field-strength.  This
calculation shows how the torsion is indeed proportional to
$F_{\mu\nu\rho}$, and not $\ep_{\mu\nu\rho\alpha\beta\gamma}
 F^{\alpha\beta\gamma}$.  A further simplification in our calculation will be
to neglect terms where derivatives land on the dilaton field.

    To begin, we expand the the terms in the Lagrangian (\ref{sezginlag})
in powers of $H_{\mu\nu\rho}$, keeping only those of quadratic or lower
order.  From (\ref{riemanntilde}) we have, up to quadratic order in
$H_{\mu\nu\rho}$,
\begin{eqnarray}
\widetilde R^{\mu\nu\rho\sigma} \widetilde R_{\mu\nu\rho\sigma}  &=&
R^{\mu\nu\rho\sigma} R_{\mu\nu\rho\sigma} - R_{\alpha\beta\mu\nu}\,
H_{\alpha\lambda\mu} H_{\beta}{}^\lambda{}_\nu +\ft12
(\nabla_\mu H_\nu{}^{\alpha\beta} - \nabla_\nu H_\mu{}^{\alpha\beta})
 \nabla^\mu H^\nu{}_{\alpha\beta} \nn\,,\\
&\rightarrow& R^{\mu\nu\rho\sigma} R_{\mu\nu\rho\sigma} -
R_{\alpha\beta\mu\nu}\,
H_{\alpha\lambda\mu} H_{\beta}{}^\lambda{}_\nu -
\ft12\nabla^\mu(\nabla_\mu H_\nu{}^{\alpha\beta} -
 \nabla_\nu H_\mu{}^{\alpha\beta})H^\nu{}_{\alpha\beta}
  \nn\,,
\end{eqnarray}
which, upon use of the Bianchi identity $dH_\3=0$ and the
zeroth-order equation of motion $\nabla^\mu H_{\mu\nu\rho}=0$ (recall we are
neglecting terms involving derivatives of $\phi$ here) gives
\begin{equation}
\widetilde R^{\mu\nu\rho\sigma} \widetilde R_{\mu\nu\rho\sigma}  =
R^{\mu\nu\rho\sigma} R_{\mu\nu\rho\sigma} - 3R^{\alpha\beta\mu\nu}\,
H_{\alpha\lambda\mu} H_{\beta}{}^\lambda{}_\nu +{\cal O}(H^3)\,.
\end{equation}
The Chern-Simons term in (\ref{sezginlag}) has the expansion
\begin{eqnarray}
{\cal L}_{CS} &=&
-\ft1{72}\beta\, \sqrt{-g} \, \ep^{\mu\nu\rho\alpha\beta\gamma}
\, \widetilde I_{\mu\nu\rho}\, H_{\alpha\beta\gamma} \nn\,,\\
&=& -\ft1{72}\beta\, \sqrt{-g} \, \ep^{\mu\nu\rho\alpha\beta\gamma}
\,\Big( I_{\mu\nu\rho}\, H_{\alpha\beta\gamma}  +3 H_{\mu\sigma\lambda}\,
 R^{\sigma\lambda}{}_{\nu\rho}\, H_{\alpha\beta\gamma}\Big) + {\cal O}(H^3)\,.
\label{csquad}
\end{eqnarray}
Using the Schoutens identity
$R^{\sigma[\lambda}{}_{\nu\rho}\, \ep^{\mu\nu\rho\alpha\beta\gamma]}
=0$, we find, after some manipulations, that the second term in
(\ref{csquad}) is proportional to the Ricci tensor and
hence by using the zeroth-order Einstein equation (with derivatives of
$\phi$ neglected), it becomes of higher than quadratic order in $H$.
To the order we are working, Lagrangian (\ref{sezginlag}) can therefore
be expanded as
\begin{eqnarray}
{\cal L} &=&\sqrt{-g} \Big[e^{-2\phi} (R-\ft1{12} H_3^2) -
   \ft1{72} \beta I_{\mu\nu\rho} H_{\alpha\beta\gamma}\,
\ep^{\mu\nu\rho\alpha\beta\gamma} -\ft34\alpha R^{\alpha\beta\mu\nu}\,
  H_{\alpha\lambda\mu} H_\beta{}^\lambda{}_\nu\Big]\,.
\end{eqnarray}

   To perform the dualisation, we next add a Lagrange multiplier term
\begin{equation}
{\cal L}_{LM} = -\ft1{36}\, \sqrt{-g}\,
\bar F_{\mu\nu\rho} H_{\alpha\beta\gamma}\, \ep^{\mu\nu\rho\alpha\beta\gamma}
\end{equation}
to the Lagrangian, where $\bar F_\3 \equiv dA_\2$,
make the changes of variables
given in (\ref{duality1}) and (\ref{duality2}), and then vary with respect to
$H_{\alpha\beta\gamma}$.  This gives the result that
\begin{equation}
F_{\mu\nu\rho} = 3\del_{[\mu} A_{\nu\rho]} - \ft12\beta I_{\mu\nu\rho}
    -\ft32\alpha R^{\alpha\beta}{}_{[\mu\nu}\, F_{\rho]\alpha\beta}\,.
\end{equation}
With $\alpha=\beta$ we see that indeed this is the correct expansion, up
to linear order in $F_\3$, of the expression
\begin{equation}
F_\3 = dA_\2 - \ft12\alpha \widehat I_3\,,
\end{equation}
where
\begin{equation}
\widehat I_3 = (\widehat \Gamma^\alpha{}_{\mu\beta} \del_\nu
\widehat\Gamma^\beta{}_{\rho\alpha} + \ft23
\widehat \Gamma^\alpha{}_{\mu\beta}
\widehat \Gamma^\beta{}_{\nu\gamma}
\widehat \Gamma^\gamma{}_{\rho\alpha})d x^\mu\wedge du^\nu\wedge dx^\rho\,,
\end{equation}
with $\widehat \Gamma^\mu{}_{\nu\rho} = \Gamma^\mu{}_{\nu\rho} -\ft12
F^\mu{}_{\nu\rho}$. Thus at the order to which we have worked here, we have
sseen how the two six-dimensional theories are related by duality.

 It should be emphasised that from (\ref{duality1}), the relationship between
the six-dimensional theory constructed in
\cite{Bergshoeff:1985mz,Bergshoeff:1986wc}, and the dimensional reduction of
the heterotic string, is non-perturbative in nature, since the sign of the
dilaton is reversed.  Consequently, the embedding of the three-dimensional
massive supergravity in string theory is also of a non-perturbative nature.
This is consistent with the fact that both
the three-dimensional and the six-dimensional theories
are complete, whereas the higher-order terms in the heterotic theory require
infinite sequences of higher curvature terms.


\section{Conclusions}


   In this paper we have obtained a new type of massive three-dimensional
gravity, by performing a 3-sphere reduction of the off-shell six-dimensional
$\cN=(1,0)$ supergravity that was constructed in
\cite{Bergshoeff:1985mz,Bergshoeff:1986wc}.  This six-dimensional
starting point is of particular interest because it is fully
supersymmetric with just quadratic curvature terms added to the
basic Poincar\'e supergravity, and hence we can obtain a closed-form
result in three dimensions. The three-dimensional theory comprises
an $\cN=1$ off-shell supergravity multiplet coupled to an on-shell scalar
multiplet which cannot be non-trivially decoupled.  The theory has three
parameters (two continuous plus one discrete).

   Because the constraints of $\cN=1$ supersymmetry in three dimensions are
weaker than those of $\cN=(1,0)$ supersymmetry in six dimensions,
we can actually relax the relations between the coefficients of the terms
in the dimensionally-reduced Lagrangian, whilst still maintaining
$\cN=1$ supersymmetry.  In this way, we then generalised our original
three-dimensional theory to one containing eight parameters (of which
seven are continuous and one is discrete).

   We also considered the possible relation between our six-dimensional
starting point in \cite{Bergshoeff:1985mz,Bergshoeff:1986wc}, and
the heterotic string in ten dimensions.  More specifically, we focussed
on the ten-dimensional supergravity considered in \cite{bergsderoo}, in
which the anomaly-cancelling $\tr(R\wedge R)$ term of the heterotic
theory was supersymmetrised.  This required the introduction of
curvature-squared terms, and in fact an infinite sequence of higher-order
curvature terms (which were not constructed in \cite{bergsderoo}).
If this theory is reduced on $T^4$ and truncated to $\cN=(1,0)$, the
resulting six-dimensional theory must evidently be the dual of the theory
constructed in \cite{Bergshoeff:1985mz,Bergshoeff:1986wc} that served as
our starting point in this paper.  This is rather remarkable, since
the reduction of \cite{bergsderoo} would yield an infinite
sequence of higher-order curvature terms, whereas the theory
in \cite{Bergshoeff:1985mz,Bergshoeff:1986wc} is exactly supersymmetric
with no curvature terms beyond the quadratic order.

   As is known from recent work, there are quite
large classes of higher-order superinvariants in three dimensions that
each involve only a finite number of terms.  So far, in six dimensions,
the only known example is the curvature-squared invariant studied in
\cite{Bergshoeff:1985mz,Bergshoeff:1986wc}.  It would be interesting
to investigate whether further such invariants might exist in six
dimensions.

   The 3-sphere reduction that we performed in this paper was a very simple
one that did not involve non-singlets under the action of the
isometry group of the sphere. It is unclear whether a consistent reduction
that retained all the $SO(4)=SU(2)_L\times SU(2)_R$
Kaluza-Klein gauge fields is possible, but
it would certainly be possible to perform a consistent DeWitt group-manifold
reduction, retaining the gauge fields of $SU(2)_R$ and all other singlets
under $SU(2)_L$.  Such reductions of six-dimensional supergravity, in
the absence of higher-order curvature terms, have been considered in the
past \cite{lupose1,lupose2,gakana}.  It would be interesting to carry
out analogous reductions including the higher-order terms.

\section*{Acknowledgement}

   We are grateful to Eric Bergshoeff, Josh Lapan and Toine van Proeyen
for discussions.
We are very grateful to the KITPC, Beijing, for hospitality during the
course of this work.
The research of C.N.P.
is supported in part by DOE grant DE-FG03-95ER40917, and the research of
E.S. is supported in part by NSF grants PHY-0555575 and PHY-0906222.

\newpage

\begin{appendix}


\section{Notation and conventions}


The six-dimensional $\Gamma$-matrices obey the Clifford algebra
$\{\Gamma_a,\Gamma_b\}=2\eta_{ab}$  with
$\eta_{ab}={\rm diag} (-,+,...,+)$\footnote{This convention
differs from that in \cite{Bergshoeff:1985mz} where
$\eta_{ab}={\rm diag}(+,+,...,+)$.
Accordingly, we let $\varepsilon^{a_1...a_6}
\to -\im \,\varepsilon^{a_1...a_6}$ in
\cite{Bergshoeff:1985mz}.
Our definition of the  Ricci tensor
$R_{\mu\nu}=g^{\lambda\tau} R_{\mu\lambda\nu\tau}$ also
differs from that of \cite{Bergshoeff:1985mz} where
$R_{\mu\nu}=g^{\lambda\tau} R_{\mu\lambda\tau\nu}$.}.
The spinors in six dimensions are symplectic Majorana-Weyl,
obeying the reality condition
\begin{equation}
(\psi_i)^* = B \Omega^{ij} \psi_j\ , \qquad i=1,2\ ,
\end{equation}
where $\Omega_{ij}=-\Omega_{ji}$ is the $Sp(1)$ invariant constant tensor,
and $B$ is constructed from six-dimensional $\Gamma$-matrices such that
\begin{equation}
\Gamma_a^* = -B\,\Gamma_a B^{-1}\ , \qquad B^T=-B\ , \qquad B^*=-B\ .
\end{equation}
The Dirac conjugate is defined as $ {\bar\psi}^i =\im\,  (\psi_i)^* \gamma_0$.
The $Sp(1)$ indices are raised and lowered using $\Omega$:
\begin{equation}
\psi^i=\Omega^{ij}\psi_j\ ,\qquad \psi_i=\psi^j\Omega_{ji}\ ,\qquad
\Omega_{ij}\Omega^{jk}=-\delta_i^k\ .
\end{equation}
The contraction of the $Sp(1)$ indices is such that
$\beps\psi=\beps^i\psi_i$.
The fermionic bilinears have the property
$\beps^i\Gamma_{a_1\cdots a_n} \psi^j= -(-1)^n
{\bar\psi}^j\Gamma_{a_n\cdots a_1} \eps^i$.

Under $SO(5,1)\to SO(2,1)\times SO(3)$, we let $a\to (a,a')$, where
now $a=0,1,2$ and $a'=1,2,3$, and the $\Gamma$-matrices decompose as
\begin{eqnarray}
\Gamma_a &=& \gamma_a\times 1\times \Sigma_1\ , \qquad
\Gamma_{a'} = 1\times \sigma_{a'} \times \Sigma_2\ ,
\nn\ww2
B &=& 1\times \sigma_2\times \Sigma_3\ ,\qquad
\Gamma_7=-\Gamma_0\Gamma_1\cdots \Gamma_6= 1\times 1\times \Sigma_3\ ,
\nn\ww2
\Gamma_{\mu_1...\mu_6} &=& -\varepsilon_{\mu_1...\mu_6} \Gamma_7\ .
\end{eqnarray}
where $\{\gamma_a,\gamma_b\}=2\eta_{ab}$ with
$\eta_{ab}={\rm diag} (-,+,+)$, and $\sigma_{a'}$ as well as $(\Sigma_1,\Sigma_2,\Sigma_3)$ are
are the standard Pauli matrices. In our conventions
$\gamma^{\phantom{\Sigma}}_{012}=1$. The supersymmetry
parameter has positive chirality $\Gamma_7\eps^i=\eps^i$, which implies
$\Sigma_3\eps^i= \eps^i$.  Our conventions for the Levi-Civita tensor
densities in six and three dimensions are that $\epsilon_{012345}=+1$ and
$\epsilon_{012}=+1$.


\section{Expansion formula}


Given the action
\begin{equation}\label{action1}
I= \int d^3 x \sqrt{-g}\, \left(c_1 R_{\mu\nu}^2 +c_2 R^2\right)\ ,
\end{equation}
its variation with respect to the metric is
\begin{equation}\label{var1}
\delta I = \int d^3 x \delta g^{\mu\nu} \sqrt{-g}\, L_{\mu\nu}\ ,
\end{equation}
where
\begin{eqnarray}\label{L}
L_{\mu\nu} &=&
-(2c_2+c_1) \nabla_\mu\nabla_\nu R +
c_1 \,\square R_{\mu\nu} +\ft12 (4c_2 + c_1) g_{\mu\nu} \,\square R
\label{fe}\ww2
&& -\ft12 (c_2 + 2c_1) g_{\mu\nu} R^2 + (2c_2+3c_1) RR_{\mu\nu}
-\ft12 c_1 \left(8R_{\mu\lambda} R^\lambda{}_\nu  - 3g_{\mu\nu}
R_{\lambda\tau}^2 \right)\ .
\end{eqnarray}
The linearizations of various quantities about AdS$_3$
with ${\bar R}=6\Lambda$, in the gauge
$\nabla^\mu H_{\mu\nu}=0$
with $H_{\mu\nu}=h_{\mu\nu}-\ft13 {\bar g}_{\mu\nu}h$, and with
total derivative terms discarded, take the form:
\begin{eqnarray}
R_{\mu\nu}^{(1)} &=& -\ft12\left(\square-6\Lambda\right) H_{\mu\nu}
-\ft16\left(\nabla_\mu\nabla_\nu+ g_{\mu\nu}\square\right) h\ ,
\nn\ww2
R^{(1)} &\equiv & \left(g^{\mu\nu} R_{\mu\nu}\right)^{(1)} = -\ft23 \left(\square+3\Lambda\right) h\ ,
\nn\ww2
G_{\mu\nu}^{(1)} &=& -\ft12 \,\square H_{\mu\nu}
-\frac16 \left(\nabla_\mu\nabla_\nu- \gb_{\mu\nu}\, \square \right) h\ ,
\nn\ww2
C_{\mu\nu}^{(1)} &=& -\ft12 \left(\square-2\Lambda\right)
\varepsilon_\mu{}^{\alpha\beta}\,
\nabla_\alpha H_{\beta\nu}\ .
\end{eqnarray}
(A superscript $^{(1)}$ indicates that the quantity on which it has been
placed is linearized around the AdS$_3$ background.  After doing this,
we then drop the bars from derivative operators in the background.)
At the higher derivative level, the following expansion formula is useful:
\begin{eqnarray}\label{hdexp}
\left(\square R_{\mu\nu}\right)^{(1)} &=& \square \left(R_{\mu\nu}^{(1)} -2\Lambda h_{\mu\nu}\right)
\nn\ww2
&=&  -\ft12 \square\left(\square-2\Lambda\right) H_{\mu\nu}
-\ft16 \nabla_\mu\nabla_\nu\left( \square+6\Lambda\right)h
-\ft16 g_{\mu\nu}\square\left(\square+2\Lambda\right)h\ .
\nn\ww2
\end{eqnarray}
Note also that $\left(\square R\right)^{(1)}= \square R^{(1)}$ and
$\left(\nabla_\mu\nabla_\nu R\right)^{(1)} = \nabla_\mu\nabla_\nu R^{(1)}$.
Other useful formulae include the arbitrary variation of the
Einstein-Hilbert Lagrangian
\begin{equation}\label{lemma}
\delta \left(\sqrt{-g} R\right) = \sqrt{-g} \left( G_{\mu\nu}
-\nabla_\mu\nabla_\nu + g_{\mu\nu} \, \square\right) \delta g^{\mu\nu}\ ,
\end{equation}
and the commutators
\begin{eqnarray}
[\square,\nabla_\mu] h &=& \nabla_\mu (\square+\Lambda) h\ ,
\nn\ww2
[\square, \nabla_\mu\nabla_\nu]h &=&
6\Lambda \left(\nabla_\mu\nabla_\nu -\ft13 \gb_{\mu\nu}\, \square \right) h\ .
\end{eqnarray}
Finally, we record the three-dimensional identity
\begin{equation}
 R_{\mu\nu}{}^{ab}=4 R_{[\mu}{}^{[a} e_{\nu]}{}^{b]}- R e_{[\mu}{}^{[a} e_{\nu]}{}^{b]}\ .
\end{equation}
%


\section{Off-shell $\cN=(1,0)$ supergravity in $6D$ in Einstein frame}


The  bosonic part of the Lagrangian is given by \cite{Bergshoeff:1985mz}
\begin{eqnarray}
e^{-1} {\cal L} &=& R +\ft14 V_\mu^{ij} V^\mu_{ij} -
\partial_\mu\phi \partial^\mu\phi  -
\ft{1}{12} e^{-2\phi} H_{\mu\nu\rho} H^{\mu\nu\rho}
\nn\ww2
&& +\frac{1}{2\times 5!} G_{\mu_1...\mu_5} G^{\mu_1...\mu_5}
-\frac{1}{5!{\sqrt 2}} \varepsilon^{\mu\nu_1...\nu_5}
G_{\nu_1...\nu_5}V_{\mu}^{ij}\delta_{ij}\,.
\end{eqnarray}
The action is invariant under off-shell supersymmetry transformations,
which, up to cubic fermions, take the form \cite{Bergshoeff:1985mz}
\begin{eqnarray}
\delta e_\mu^a &=& \ft12 \beps\, \Gamma^a\psi_\mu\ ,
\nn\ww2
\delta \psi_\mu &=&
{\cal D}_\mu(\omega)\eps -\ft{1}{48} e^{-\phi} \Gamma\cdot H
\Gamma_\mu\epsilon +\Gamma_\mu\eta\ ,
\nn\ww2
\delta B_{\mu\nu} &=& -\beps\, \Gamma_{[\mu}\psi_{\nu]} e^\phi -\beps
\Gamma_{\mu\nu}\psi\ ,
\nn\ww2
\delta V_\mu^{ij} &=& -4\beps^{(i}\, \phi^{j)}_\mu -
\ft{1}{12} e^{-2\phi} \beps^{(i}\,  \Gamma_\mu \Gamma\cdot H \psi^{j)}
-2e^{-\phi}\beps^{(i}\, {\widehat{\slashed{D}}} \psi^{j)}
-4{\bar\eta}^{(i}\,\psi^{j)}_\mu\ ,
\nn\ww2
\delta C_{\mu\nu\rho\sigma} &=&   2{ \sqrt 2}\, \beps^i\,
\Gamma_{[\mu\nu\rho}  \psi^j_{\sigma]}\, \delta_{ij}\ ,
\nn\ww2
\delta \psi &=& \ft14 e^{\phi} \Gamma^\mu \partial_\mu \phi \eps
  -\ft{1}{48} \Gamma\cdot H \eps  -e^\phi \eta\ ,
\nn\ww2
\delta\phi &=& e^{-\phi} \beps\psi\ ,
\end{eqnarray}
where \cite{Bergshoeff:1985mz}
\begin{eqnarray}
\phi_\mu &=& -\ft{1}{16}\left(\Gamma^{\rho\sigma}\Gamma_\mu
-\ft35\Gamma_\mu\Gamma^{\rho\sigma}\right) {\psi'}_{\rho\sigma}
-\ft{1}{60}\Gamma_\mu \chi\ ,
\nn\ww2
\chi&=& 6e^{-\phi}\Gamma^\mu
{\widehat \cD}_\mu \psi+\ft14 e^{-2\phi} \cdh\psi\ ,
\nn\ww2
\eta_i &=&
\Big( -\frac1{4\sqrt2}  \Gamma^\mu V_\mu^{(j}{}_k\,
 \delta^{\ell)k}\,\epsilon_j +
\frac{1}{5!\times 8{\sqrt 2}}
  \Gamma^{\mu_1...\mu_5} G_{\mu_1...\mu_5}\eps^\ell\Big)
\delta_{\ell i}\ ,
\label{eta2}
\end{eqnarray}
and
\bea
\widehat{\cD}_\mu\psi &=& \cD_\mu(\omega)\psi  +\ft{1}{48} \cdh\psi_\mu
-\ft14 e^\phi \Gamma^\nu \partial_\nu \phi \psi_\mu+e^\phi \phi_\mu\ ,\\
\psi'_{\mu\nu} &=& 2 \cD_{[\mu}(\omega)\psi_{\nu]} + \ft1{24} e^{-\phi}\,
  \Gamma\cdot H\, \Gamma_{[\mu}\, \psi_{\nu]}\, .
\eea

It was observed in \cite{Bergshoeff:1986wc} that the fields $\phi$ and
$\psi$ can be eliminated from the transformation rules of the multiplet
of fields $(e_\mu,\psi_\mu, V_\mu^i,B_{\mu\nu})$ by means of the
the field redefinitions \eq{redefs}. These redefinitions lead to drastic
simplifications and, dropping the hats on ${\widehat e}_\mu^a,
{\widehat\psi}_\mu$ and ${\widehat\eps}$ for simplicity, yield
the transformation rules \eq{susy6}.

\end{appendix}

\end{document}